\documentclass[11pt]{article}
\usepackage{latexsym}
\usepackage{graphics,graphicx}
\usepackage{amsmath,amsfonts,amssymb}
\usepackage{esint}
\usepackage{epsfig}
\usepackage{lscape}
\usepackage{hyperref}
\def\NO{\nonumber}

\newcommand{\be}{\begin{equation}}
\newcommand{\ee}{\end{equation}}

\def\bea{\begin{eqnarray}}
\def\eea{\end{eqnarray}}

\def\beqx{\begin{displaymath}}
\def\eeqx{\end{displaymath}}

\newcommand{\bmat}{\left(\begin{array}}
\newcommand{\emat}{\end{array}\right)}

\def\a{\alpha}
\def\b{\beta}
\def\c{\chi}
\def\d{\delta}
\def\e{\epsilon}

\def\g{\gamma}

\def\k{\kappa}
\def\l{\lambda}
\def\m{\mu}
\def\n{\nu}

    \def\om{\omega}
\def\p{\pi}

    \def\th{\theta}
\def\r{\rho}
\def\s{\sigma}

\def\x{\xi}
\def\z{\zeta}
\def\D{\Delta}

\def\G{\Gamma}

    \def\Om{\Omega}
\def\P{\Pi}

    \def\Th{\Theta}
\def\S{\Sigma}

\def\X{\Xi}

\def\vf{\varphi}

\def\ca{{\cal A}}

\def\cf{{\cal F}}

\def\ch{{\cal H}}

\def\cl{{\cal L}}
\def\cm{{\cal M}}
\def\cn{{\cal N}}
\def\co{{\cal O}}
\def\cp{{\cal P}}

\def\car{{\cal R}}
\def\cs{{\cal S}}
\def\ct{{\cal T}}
\def\cu{{\cal U}}

\def\bo{{\raise-.3ex\hbox{\large$\Box$}}}               
\def\pa{\partial}                                       
\def\face{{\raise.2ex\hbox{$\displaystyle \bigodot$}\mskip-2.2mu \llap {$\ddot
        \smile$}}}                                   
\def\>{\rangle}                                      
\def\<{\langle}                                      

\newcommand{\sub}[1]{\phantom{}_{(#1)}\phantom{}}    
\def\lbar#1{\ensuremath{\overline{#1}}}              
\def\leftrightarrowfill{$\mathsurround=0pt \mathord\leftarrow \mkern-6mu
        \cleaders\hbox{$\mkern-2mu \mathord- \mkern-2mu$}\hfill
        \mkern-6mu \mathord\rightarrow$}        
\def\dvec#1{\vbox{\ialign{##\crcr
        \leftrightarrowfill\crcr\noalign{\kern-1pt\nointerlineskip}
        $\hfil\displaystyle{#1}\hfil$\crcr}}}           
\def\Tr{{\rm Tr \,}}                                    

\def\-{\hphantom{-}}

\newlength{\ytwd}
\newlength{\ytht}

\title{
\begin{minipage}{6.5in}
\begin{flushright}
 {\small CERN-PH-TH/2011-143}
\end{flushright}
\end{minipage}\\
\vskip 1.0in
Holographic Renormalization of general dilaton-axion gravity
}

\author{Ioannis Papadimitriou\\ \\
{\it Department of Physics, CERN -- Theory Division,}\\ \\ {\it CH--1211 Geneva 23, Switzerland}\\ \\
e-mail: {\tt Ioannis.Papadimitriou@cern.ch}}

\date{}

\begin{document}

\maketitle

\begin{abstract}

We consider a very general dilaton-axion system coupled to Einstein-Hilbert gravity in arbitrary dimension and we carry out
holographic renormalization for any dimension up to and including five dimensions. This is achieved by developing 
a new systematic algorithm for iteratively solving the radial Hamilton-Jacobi equation in a derivative expansion. The boundary 
term derived is valid not only for asymptotically AdS backgrounds, but also for more general asymptotics, including non-conformal 
branes and Improved Holographic QCD. In the second half of the paper, we apply the general result to Improved Holographic QCD with
arbitrary dilaton potential. In particular, we derive the generalized Fefferman-Graham asymptotic expansions and provide  
a proof of the holographic Ward identities.      

\end{abstract}

\newpage

\tableofcontents
\addtocontents{toc}{\protect\setcounter{tocdepth}{2}}
\renewcommand{\theequation}{\arabic{section}.\arabic{equation}}

\section{Introduction}
\setcounter{equation}{0}

Understanding the holographic dictionary for holographic models in non asymptotically AdS spaces has been a long standing
problem. It has been a pressing question ever since physically promising holographic dualities involving 
non asymptotically AdS backgrounds, such as the Klebanov-Strassler \cite{Klebanov:2000hb} and Maldacena-N\'u\~nez 
\cite{Maldacena:2000yy} backgrounds, were found, but it has become even more relevant with the recent interest 
in the phenomenological application of holography to condensed matter physics and models of QCD. Even though 
numerous attempts have been made to understand aspects of the dictionary of some of these systems, there are very few cases 
where a systematic and extensive understanding has been achieved for non asymptotically AdS backgrounds. These 
include the analyses of the dictionary for non-conformal branes \cite{Wiseman:2008qa,Kanitscheider:2008kd} and 
for Schr\"odinger backgrounds \cite{Guica:2010sw}.        

However, it cannot be overemphasized that the process of understanding the holographic dictionary, in any holographic model and
any background, even beyond the supergravity approximation, can be split into two conceptually distinct steps. The first step 
is intrinsically related with the ``bulk'' holographic model. Namely, one must identify a suitable boundary in 
the bulk theory and construct a reduced phase space of the theory in terms of data on that boundary. This step 
is exactly analogous to the Fefferman-Graham reconstruction of the bulk geometry in asymptotically hyperbolic 
manifolds from boundary data \cite{FG}. A systematic way of addressing this question and its connection to a certain 
variational problem at infinity in the most general setting was discussed in \cite{Papadimitriou:2010as}. The approach 
developed in \cite{Papadimitriou:2010as} in principle allows one to algorithmically construct this reduced phase space 
for any bulk model. Having completed this step, one not only has achieved a reformulation of the bulk dynamics in terms 
of a symplectic space of boundary data that can unambiguously be identified with the symplectic space of renormalized 
observables in any holographically dual theory, but also has automatically made the variational problem of the 
bulk theory well defined, which implies that the on-shell action is finite \cite{Papadimitriou:2005ii,Papadimitriou:2010as}.

Only once this first step has been completed, one can directly compare the symplectic space of boundary data with 
the symplectic space of gauge-invariant observables in any candidate holographic dual. This mapping is simply 
the classical version of the Hilbert space isomorphism one expects in a fully quantum mechanical holographic duality. 
On the bulk side one has the symplectic space of a classical system, being that classical strings or classical gravity, 
while on the field theory side the Hilbert space reduces to a classical symplectic space in some limit where the number of 
degrees of freedom becomes infinite \cite{ElShowk:2011ag}.       

In this paper we consider a generic dilaton-axion system coupled to Einstein-Hilbert gravity in arbitrary dimension
with the action (\ref{action-scalars}). This system contains the standard dilaton-axion system in AdS$_5$ dual to the complexified 
coupling of $\cn=4$ super Yang-Mills in four dimensions, non-conformal branes \cite{Wiseman:2008qa,Kanitscheider:2008kd}, 
as well as Improved Holographic QCD \cite{Gursoy:2007cb} as special cases. The last two examples admit non asymptotically AdS vacua, 
and so the standard dictionary for asymptotically AdS gravity is not applicable. Our aim here will be to carry out this two-step 
procedure outlined above for this general dilaton-axion system and to explore in more detail the consequences for the 
model of Improved Holographic QCD.   

The semi-phenomenological holographic model dual to large $N_c$ Yang-Mills theory put forward in \cite{Gursoy:2007cb}
(see \cite{Gursoy:2010fj} for an extensive review) is based
on the five dimensional two-derivative (Euclidean) bosonic supergravity action
\be\label{log-action}
S=-M_{pl}^3N_c^2\int d^{5}x \sqrt{g}\left(R[g]-\x^2\l^{-2}
\pa_\m \l\pa^\m\l-Z(\l)\pa_\m \c\pa^\m \c+V(\l)\right),
\ee
where the Planck mass $M_{pl}^3=1/g_s^2\ell_s^3$ is related to the five dimensional Newton's constant by
$(16\pi G_{5} )^{-1}=M_{pl}^3N_c^2$. The field content of this action consists of the five dimensional
metric $g_{\m\n}$, a dilaton $\l$, and an axion $\c$. These are respectively designed to
describe the dynamics of the lowest dimension gauge-invariant operators of pure Yang-Mills theory, namely
the stress tensor, $T_{ij}$, $\Tr(F^2)$ and $\Tr(F\widetilde{F})$. In particular, $\l$ is proportional to the
't Hooft coupling, $N_c g_{YM}^2$, while $\c$ is related to the instanton angle $\th_{YM}$. The constants
of proportionality are not known a priori, but can be determined by comparing the perturbative UV expansion of
the beta functions for the 't Hooft coupling and $\th_{YM}$ with the corresponding holographic beta functions
for the bulk fields $\l$ and $\c$ respectively. These proportionality constants are nevertheless scheme
dependent and so they do not affect the value of any physical observable \cite{Gursoy:2010fj}.

The holographic model is defined by the potential $V(\l)$ and the function $Z(\l)$, as well as the constant $\x\neq 0$,
corresponding to the normalization of the kinetic term of the dilaton.\footnote{In \cite{Gursoy:2007cb} $\x^2=4/3$, but
here we prefer to keep it arbitrary.} Anticipating that such a model could possibly originate in a non-critical
string theory in five dimensions \cite{Gursoy:2007cb}, the metric and the dilaton are expected to come from the
NSNS sector while the axion comes from the RR sector. As was argued in \cite{Gursoy:2007cb}, this implies that the
kinetic term of the axion should be $\co(1/N_c^2)$ relative to the Einstein-Hilbert term, the dilaton kinetic term, and
the scalar potential. Hence, $Z(\l)=\co(1/N_c^2)$, while $\x$ and $V(\l)$ are $\co(N_c^0)$.

The form of the functions $Z(\l)$ and $V(\l)$ can be constrained by physical input. Firstly, asymptotic freedom means
that at the UV the theory is conformal and so the dual string vacuum should be asymptotically AdS$_5$. Of course, it
also means that the theory is weakly coupled at the far UV, and so a two-derivative gravity approximation of the form
(\ref{log-action}) cannot a priori be trusted in that limit. However, the assumption in \cite{Gursoy:2007cb} is that
the effect of the higher derivative terms affecting the dynamics of the lowest lying fields can somehow be incorporated
into a an effective cosmological constant in the potential $V(\l)$, thus allowing the two-derivative action
(\ref{log-action}) to admit asymptotically AdS solutions. Although this argument is admittedly difficult, if not impossible,
to defend from a string theoretic point of view, the attitude in \cite{Gursoy:2007cb} is that this assumption can be a posteriori
justified by the success of the model in reasonably describing various qualitative properties of Yang-Mills theory. Accepting 
this assumption, means that the dilaton $\l$ must vanish in the far UV and the functions $V(\l)$ and $Z(\l)$ must admit Taylor
expansions of the form\footnote{Potentials that contain non-analytic terms at order higher than $\co(\l^2)$ have been 
considered in the literature. The analysis below applies to such more general potentials as well.}
\be\label{Taylor-5}
V(\l)=\frac{12}{ \ell^2}\left(1+\sum_{n=1}^{\infty}V_n\l^n\right),\qquad
Z(\l)=(M_{pl}^3N_c^2)^{-1}\sum_{n=0}^{\infty}Z_n\l^n,
\ee
where $\ell$ is the radius of the AdS corresponding to the UV fixed point, and $V_n$ and $Z_n$ are $\co(N_c^0)$. The
original motivation for these expansions was that they should be the holographic analogue of the perturbative expansions 
of the beta functions $\b_\l$ and $\b_\c$ of the 't Hooft coupling and instanton angle respectively \cite{Gursoy:2007cb}.
In particular, the coefficients $V_1$ and $V_2$ were argued to be related respectively to the one- and two-loop beta function 
of the 't Hooft coupling. In order to account for the logarithmic running of the coupling, therefore, one demands 
that $V_1\neq0$. As we shall see below, the requirement that $V_1\neq0$ is indeed crucial for the asymptotic solutions 
to logarithmically deviate from strictly AdS asymptotics. However, the beta function is a scheme-dependent quantity 
which does not correspond to any physical observable. As we shall see below, the holographic Ward identities only involve 
physical, renormalization group invariant quantities. 

Further constraints on the functions $V(\l)$ and $Z(\l)$ are imposed by the IR properties of the model, i.e. as
$\l\to\infty$. Specifically, confinement and the absence of certain `bad' singularities in the IR require
that \cite{Gursoy:2007cb}
\be
V(\l)\sim \l^{2Q}\left(\log\l\right)^P,\,\, {\rm as}\,\, \l\to\infty, \,\,{\rm with}\,
\left\{\begin{matrix}
      2/3<Q<2\sqrt{2}/3, & P\,\, {\rm arbitrary},\\
      Q=2/3, & P\geq 0.
      \end{matrix}\right.
\ee
Moreover, requiring an asymptotically linear glueball spectrum, $m_n^2\sim n$, uniquely picks out $Q=2/3$, $P=1/2$.
Certain conditions for the strongly coupled limit of $Z(\l)$ are also necessary in some cases \cite{Gursoy:2007cb}.
Here we will keep the functions $V(\l)$ and $Z(\l)$ completely general, however, assuming only
that\footnote{We will consider an arbitrary boundary dimension $d$ from now on, in order to maintain a wider applicability 
of our results.}
\be\label{generic-potential}
V(\l)=\frac{d(d-1)}{\ell^2}\left(1+V_1\l+V_2\l^2\right)+\widetilde{V}(\l),\quad V_1\neq0,\,\widetilde{V}(\l)=o(\l^2),\, {\rm as}\, \l\to0.
\ee
In particular, we allow $\widetilde{V}(\l)$ to contain non analytic terms provided they are subleading compared to $\l^2$ as
$\l\to0$. 

Given the functions $V(\l)$ and $Z(\l)$, the Yang-Mills vacuum is described by extrema of (\ref{log-action})
with four-dimensional Poincar\'e invariance. Such solutions are domain walls of the form
\be
ds^2=dr^2+e^{2A(r)}dx^idx^i,\quad \l=\l(r),\quad \c=\c(r).
\ee
Such backgrounds are extrema of (\ref{log-action}) provided they satisfy
\bea\label{dw_eqs}
&&\dot{A}^2-\frac{1}{d(d-1)}\left(\x^2\l^{-2}\dot{\l}^2+Z(\l)\dot{\c}^2+V(\l)\right)=0,\NO\\
&&\ddot{A}+d\dot{A}^2-\frac{1}{d-1}V(\l)=0,\NO\\
&&2\x^2\l^{-2}\left(\ddot{\l}+d\dot{A}\dot{\l}-\l^{-1}\dot{\l}^2\right)-Z'(\l)\dot{\c}^2+V'(\l)=0,\NO\\
&&Z(\l)\ddot{\c}+Z'(\l)\dot{\l}\dot{\c}+d Z(\l)\dot{A}\dot{\c}=0,
\eea
where $\dot{}$ denotes differentiation w.r.t. the radial coordinate $r$. These equations are automatically solved
provided $A$, $\l$ and $\c$ satisfy the first order flow equations
\be\label{flow_eqs}
\dot{A}=-\frac{1}{d-1}W(\l,\c),\quad \dot{\l}=\x^{-2}\l^2\frac{\pa W(\l,\c)}{\pa\l},\quad
\dot{\c}=Z^{-1}(\l)\frac{\pa W(\l,\c)}{\pa \c},
\ee
where the `superpotential' $W(\l,\c)$ is determined by the scalar potential $V(\l)$ and the function $Z(\l)$
via the equation
\be\label{potential_eq}
V(\l)=-\x^{-2}\l^2\left(\frac{\pa W}{\pa\l}\right)^2-Z^{-1}(\l)
\left(\frac{\pa W}{\pa \c}\right)^2+\frac{d}{d-1}W^2.
\ee
In particular, any solution $W(\l,\c)$ of (\ref{potential_eq}) uniquely specifies a Yang-Mills vacuum. Notice
that the last equation in (\ref{dw_eqs}) can be integrated exactly to obtain
\be
\dot\c=cZ^{-1}(\l)e^{-dA},
\ee
where $c$ is an integration constant. From the last equation in (\ref{flow_eqs}) then follows that
\be\label{integrated-axion}
\frac{\pa W(\l,\c)}{\pa\c}=ce^{-dA}.
\ee
As we will see later, this result has very significant consequences for the form of the counterterms required to make
the variational problem for the action (\ref{log-action}) well defined. As far as possible vacuum solutions are
concerned, this relation means that there are two broad classes of vacua depending on whether $c$ is zero or not. Namely,
if $c=0$, then $W$ is independent of the axion and so the axion is just a constant in the background, corresponding to the
value of the instanton angle $\th_{YM}$. Vacua with $c\neq 0$ allow for a non-trivial axion profile, corresponding to
giving a VEV to the operator $\Tr(F\widetilde{F})$. In particular, using the result (\ref{axion-1pt}) below, and assuming
$\co_\c=\Tr (F\tilde F)$, we obtain $\langle\Tr(F\widetilde{F})\rangle=c/\k^2$. Additionally, generic
vacua are classified according to whether the operator dual to the dilaton acquires a vacuum expectation value. As follows
from the exact one-point function (\ref{dilaton-1pt}), this happens iff the UV expansion of the superpotential $W(\l,\c)$ 
contains a non-zero term of the form $e^{-dA}\sim\l^{d\n}\exp(-d/b_0\l)$, where 
\be
\n=\frac{2\x^2}{d-1}-\frac1d-\frac{(d-1)V_2}{\x^2b_0^2}.
\ee
Poincar\'e invariant vacua are, therefore, parameterized by {\em three} constants: the dilaton and axion VEVs, as well 
as the instanton angle $\th_{YM}$. As we shall see, the source of the dilaton is a gauge freedom and
its value can be thought of as the energy scale. In particular, it's value is not a physical observable and 
does not correspond to a coordinate in the moduli space of vacua. This is in good agreement with what one expects 
from a holographic model describing pure Yang-Mills theory.    

The rest of the paper is organized as follows. In Section \ref{general-boundary-term} we define the general
dilaton-axion gravity theory we will consider and we formulate its dynamics in terms of a radial Hamiltonian.
We then develop a systematic iterative procedure for solving the Hamilton-Jacobi equation in a derivative expansion 
for this general class of theories. Using this procedure the full boundary term that makes the variational problem 
at infinity well defined is derived for spacetime dimension up to and including five dimensions. This result, 
summarized in Tables \ref{order-1} and \ref{order-2}, is the main result of the paper. In Section \ref{IHQCD-application}
we apply this general result to IHQCD. After writing down explicitly the boundary counterterms for IHQCD, we 
systematically derive the generalized Fefferman-Graham asymptotic expansions and give explicit expressions 
for the exact renormalized one-point functions in the presence of sources for the stress tensor, dilaton and 
axion operators. These are given in equations (\ref{stress-tensor-1pt}), (\ref{dilaton-1pt}) and (\ref{axion-1pt}) 
respectively. The section ends with a detailed discussion of the asymptotic bulk diffeomorphisms that preserve
the form of the asymptotic expansions, leading to a proof of the holographic Ward identities. Some concluding 
remarks follow in Section \ref{conclusions}, while some technical details of the main calculation are presented in
Appendix \ref{functional-integration}. Finally, in Appendix \ref{constant-potential} we apply the main result 
to the case of a constant dilaton potential, and derive general expressions for the exact one-point functions in
the presence of sources for the fully coupled stress tensor--dilaton--axion sector of $\cn=4$ super Yang Mills in four
dimensions. To our knowledge this is the first time that the general form of these one-point functions has been 
derived.

\section{Boundary term for generic dilaton-axion gravity}
\label{general-boundary-term}
\setcounter{equation}{0}

Since the action (\ref{log-action}) already contains essentially arbitrary functions of the dilaton, deriving the 
appropriate boundary term that makes its variational problem well defined is not much easier than 
considering instead the slightly more general action
\be\label{action-scalars}
S=-\frac{1}{2\k^2} \left(\int_\cm d^{d+1}x\sqrt{g}\left(R[g]
-\pa_\m\vf\pa^\m\vf-Z(\vf)\pa_\m\chi\pa^\m\chi+V(\vf)\right)+\int_{\pa\cm}d^dx\sqrt{\g}2K\right),
\ee
where the spacetime dimension is taken to be arbitrary and we have introduced a canonical dilaton field, $\vf$,
related to the dilaton $\l$ in (\ref{log-action}) by $\vf=\x\log \l$. Moreover, we have added the standard Gibbons-Hawking 
term \cite{Gibbons:1976ue} and the constant $\k$ is related to Newton's constant in $d+1$ dimensions by $\k^2=8\p G_{d+1}$. 

Note that this action contains as special cases a very large class of theories considered in the literature. Apart, 
from IHQCD, other special cases include the standard dilaton-axion of $\cn=4$ super Yang-Mills\footnote{See Appendix 
\ref{constant-potential}
for the explicit results in this case and \cite{Mateos:2011ix,Mateos:2011tv}, where these results have been recently used.}, 
as well as non-conformal branes \cite{Wiseman:2008qa,Kanitscheider:2008kd}. By deriving the appropriate boundary term for the 
action (\ref{action-scalars}), therefore, we automatically carry out the holographic renormalization of all these theories, 
whether asymptotically AdS or not, and without the need of any field redefinition. Explicit results for the case of 
non-conformal branes will be presented elsewhere \cite{Papadimitriou:non-conformal-branes}.

\subsection{Hamiltonian formulation of the variational problem}

The variational problem at infinity for the action (\ref{action-scalars}) is not well defined as it stands 
\cite{Papadimitriou:2005ii,Papadimitriou:2010as}. For the case of asymptotically AdS gravity it was first shown in 
\cite{de_Boer:1999xf} that a certain asymptotic solution
of the radial Hamilton-Jacobi equation renders the on-shell action finite. It was later shown \cite{Papadimitriou:2004ap} that 
the boundary term obtained by solving the radial Hamilton-Jacobi solution is the same as the one obtained via the standard method 
of holographic renormalization \cite{Henningson:1998gx,Balasubramanian:1999re,Emparan:1999pm,Kraus:1999di,de_Boer:1999xf,
de_Haro:2000xn,Bianchi:2001kw,Martelli:2002sp}. In \cite{Papadimitriou:2005ii} it was pointed out that the boundary 
term required to make the on-shell action finite is in fact the same boundary term required to render the variational 
problem at infinity well defined, while in \cite{Papadimitriou:2010as} it was pointed out that this conclusion holds more generally, 
not just for non-asymptotically AdS gravity, but also for non-gravitational variational problems with a boundary at infinity, 
{\em provided} the variations are confined within a space of asymptotic solutions carrying a well defined symplectic form. Moreover,
this boundary term always corresponds to a solution of the radial Hamilton-Jacobi equation. 
In practical terms this means that the leading solution of the Hamilton-Jacobi equation determined from the leading asymptotic 
form of the solutions {\em must not contain transverse derivatives} \cite{Papadimitriou:2010as}. If this is not the case, it simply 
means that the space of asymptotic solutions corresponding to the chosen leading asymptotics is not well defined and it does not 
carry a suitable symplectic form. In such cases, one must first perform some kind of `Kaluza-Klein reduction' to trivialize the 
transverse derivatives appearing in the leading radial asymptotics, derive an effective action for the KK fields, and
then solve the radial Hamilton-Jacobi equation for {\em this} effective action of the KK fields \cite{Papadimitriou:2010as}. 
The condition that the leading asymptotic form of the solutions should give rise to a solution of the Hamilton-Jacobi 
equation whose leading asymptotic form does not contain any transverse derivatives is automatically satisfied in the case
of IHQCD, asymptotically AdS gravity, or non-conformal branes and we will assume that it is the case in the analysis below.   

To formulate the variational problem we start by writing the metric
in an ADM decomposition \cite{Arnowitt:1960es}, but with Hamiltonian time being replaced by the radial coordinate, $r$, emanating
from the boundary at infinity. Namely, we write
\be\label{ADM-metric}
ds^2=(N^2+N_iN^i)dr^2+2N_idrdx^i+\g_{ij}dx^idx^j,
\ee
where $N$ and  $N_i$ are respectively the lapse and shift functions,
and $\g_{ij}$ is the induced metric on the hypersurfaces $\S_r$
of constant radial coordinate $r$.  The metric $g_{\m\n}$ is therefore
replaced in the Hamiltonian description by the three fields $\{N,N_i,\g_{ij}\}$ on $\S_r$.
In terms of these variables the Ricci scalar takes the form
\be
R[g]=R[\g]+K^2-K_{ij}K^{ij}+\nabla_\m (-2Kn^\m+2n^\r\nabla_\r n^\m),
\ee
where $R[\g]$ is the Ricci scalar of the induced metric $\g_{ij}$, the extrinsic curvature, $K_{ij}$, of the hypersurface
$\S_r$ is given by
\be\label{extr-curv}
K_{ij}=\frac{1}{2N}\left(\dot{\g}_{ij}-D_iN_j-D_jN_i\right),
\ee
and $D_i$ is the covariant derivative w.r.t. the induced metric $\g_{ij}$. Moreover, $K=\g^{ij} K_{ij}$ and
$n^\m=\left(1/N,-N^i/N\right)$, is the outward unit normal vector to $\S_r$. The total derivative term
in this decomposition of the bulk Ricci scalar is an indication of the need for the Gibbons-Hawking term. Evaluating
this term on $\S_r$ we see that it gives a contribution which is precisely canceled by the Gibbons-Hawking term. We
therefore arrive at a Lagrangian description of the dynamics of the induced fields $\{N,N_i,\g_{ij}\}$ on $\S_r$, namely
\bea\label{lagrangian-gf-scalars}
2\k^2L&=&-\int_{\S_r}d^dx\sqrt{\g}N\left(R[\g]+K^2-K^i_jK^j_i\right)\NO\\
&&+\int_{\S_r}d^dx\sqrt{\g}N\left\{\frac{1}{N^2}(\dot{\vf}^2+Z(\vf)\dot{\chi}^2)
-\frac{2N^i}{N^2}(\dot{\vf}\pa_i\vf+Z(\vf)\dot{\chi}\pa_i\chi)
\right.\NO\\
&&\left.
+\left(\g^{ij}+\frac{N^iN^j}{N^2}\right)(\pa_i\vf\pa_j\vf+Z(\vf)\pa_i\chi\pa_j\chi)-V(\vf)\right\}.
\eea
Note that this Lagrangian involves no kinetic terms for the fields $N$ and $N_i$, which are therefore Lagrange multipliers, leading
to constraints. The canonical momenta conjugate to $\g_{ij}$, $\vf$ and $\chi$ are now obtained respectively as
\bea\label{momenta-scalars}
&&\pi^{ij} \equiv\frac{\d L}{\d\dot\g_{ij}}= -\frac{1}{2\k^2}\sqrt{\g} \left(K\g^{ij}-K^{ij}\right), \NO\\
&&\p_\vf\equiv\frac{\d L}{\d\dot\vf}= \frac{1}{\k^2N}\sqrt{\g} \left(\dot{\vf}-N^i\pa_i\vf\right), \NO\\
&&\p_\chi\equiv\frac{\d L}{\d\dot\c}= \frac{1}{\k^2N}\sqrt{\g} Z(\vf)\left(\dot{\chi}-N^i\pa_i\chi\right),
\eea
while the momenta conjugate to $N$ and $N_i$ vanish identically. The Hamiltonian is given by
\be\label{hamiltonian-scalars}
H=\int_{\S_r} d^dx \left(\p^{ij}\dot{\g}_{ij}+\p_\vf\dot{\vf}+\p_\chi\dot{\chi}\right)-L=
\int_{\S_r} d^dx\left(N\ch+N_i\ch^i\right),
\ee
where
\bea\label{constraints-scalars}
\ch&=&  2\k^2\g^{-\frac12}\left(\p^i_j\p^j_i-\frac{1}{d-1}\p^2+\frac14\p_\vf^2+\frac14Z^{-1}(\vf)\p_\chi^2\right)\NO\\
&&+\frac{1}{2\k^2}\sqrt{\g}\left(R[\g]-\pa_i\vf\pa^i\vf-Z(\vf)\pa_i\chi\pa^i\chi+V(\vf)\right),\NO\\
\ch^i&=& -2D_j\p^{ij}+\p_\vf\pa^i\vf+\p_\chi\pa^i\chi.
\eea
Hamilton's equations for the fields $N$ and $N_i$ lead respectively to the Hamiltonian and momentum constraints
\be
\label{constraints}
\ch=0,\quad \ch^i=0.
\ee
Moreover, the symplectic form is given by
\be\label{symplectic-form}
\Om=\int_{\S_r}d^dx \left(\d\p^{ij}\wedge \d\g_{ij}+\d\p_\vf\wedge \d\vf+\d\p_\c\wedge \d\c\right),
\ee
and it is independent of the value of the radial coordinate $r$ \cite{Lee:1990nz}.

The variational problem for the action (\ref{action-scalars}) can now be formulated in a regularized space, $\cm_{r_o}$,
whose boundary is defined as the surface $\S_{r_o}$ for some fixed $r_o$. Provided $r_o$ is sufficiently large, $\S_{r_o}$
is diffeomorphic to the boundary $\pa\cm$ at infinity. Adding then a generic boundary term, $S_b$, to the action
(\ref{action-scalars}) defined on $\cm_{r_o}$ and considering a generic variation we obtain \cite{Papadimitriou:2010as}
\bea\label{action-variation}
\d(S+S_b)&=&\int_{\cm_{r_o}} d^{d+1}x(EOMs)+\left.(L+\dot{S}_b)\right|_{r_o}\d r_o\NO\\
&&+\int_{\S_{r_o}}d^dx \left(\left(\p^{ij}+\frac{\d S_b}{\d\g_{ij}}\right)\d\g_{ij}
+\left(\p_\vf+\frac{\d S_b}{\d\vf}\right) \d\vf
+\left(\p_\c+\frac{\d S_b}{\d\c}\right)\d\c\right),
\eea
where the integrand of the integral over the bulk of $\cm_{r_o}$ is proportional to the equations of motion. In order for
the variational problem at $r_o\to\infty$ to be well defined with boundary conditions imposed on the induced fields
$\g_{ij}$, $\vf$ and $\c$, i.e. for a generic variation of the action with such boundary conditions to imply the
equations of motion, one must demand that \cite{Papadimitriou:2010as}
\be
\left.(L+\dot{S}_b)\right|_{r_o}\xrightarrow{r_o\to\infty} 0.
\ee
The variational problem is then defined by considering variations of $\g_{ij}$, $\vf$ and $\c$ within the space
of generic asymptotic solutions of the equations of motion so that
\be
\left.S_b\right|_{r_o}=-\left.\cs\right|_{r_o},
\ee
where $\cs$ is Hamilton's principal functional, i.e. a solution of the Hamilton-Jacobi equation where
the values of the induced fields $\g_{ij}$, $\vf$ and $\c$ on $\S_{r_o}$ are totally arbitrary. It follows from
(\ref{action-variation}) that
\bea\label{HJ-momenta}
\left.\p^{ij}\right|_{r_o}=\left.\frac{\d\cs}{\d\g_{ij}}\right|_{r_o},\quad
\left.\p_\vf\right|_{r_o}=\left.\frac{\d\cs}{\d\vf}\right|_{r_o},\quad
\left.\p_\c\right|_{r_o}=\left.\frac{\d\cs}{\d\c}\right|_{r_o},
\eea
where $\cs$ is identified with the on-shell value of the action $S$ on solutions with arbitrary boundary values
for $\g_{ij}$, $\vf$ and $\c$ on $\S_{r_o}$. Including the boundary term $S_b$ in these relations leads to
the canonically transformed momenta \cite{Papadimitriou:2010as}
\bea\label{HJ-momenta-ren}
\left.\P^{ij}\right|_{r_o}=\left.\frac{\d(\cs+S_b)}{\d\g_{ij}}\right|_{r_o},\quad
\left.\P_\vf\right|_{r_o}=\left.\frac{\d(\cs+S_b)}{\d\vf}\right|_{r_o},\quad
\left.\P_\c\right|_{r_o}=\left.\frac{\d(\cs+S_b)}{\d\c}\right|_{r_o}.
\eea

In order to determine the boundary term $S_b$ we need to determine the asymptotic form of Hamilton's
principal functional. In other words we need to solve the Hamilton-Jacobi equation in a certain asymptotic sense, which
we will specify below. The Hamilton-Jacobi equation can be derived from the following simple argument. Let
$\cs_{r_o}\equiv \left.\cs\right|_{r_o}$ denote the on-shell action $S$ with arbitrary boundary values
for $\g_{ij}$, $\vf$ and $\c$ on $\S_{r_o}$. Then,
\bea
\dot{\cs}_{r_o}&=&\frac{\pa\cs_{r_o}}{\pa r_o}+\int_{\S_{r_o}} d^dx\left(\dot{\g}_{ij}[\g,\vf,\c]\frac{\d}{\d\g_{ij}}+
\dot{\vf}[\g,\vf,\c]\frac{\d}{\d\vf}+\dot{\c}[\g,\vf,\c]\frac{\d}{\d\c}\right)\cs_{r_o}\NO\\
&\stackrel{(\ref{HJ-momenta})}{=}&\frac{\pa\cs_{r_o}}{\pa r_o}+\int_{\S_{r_o}} d^dx \left(\p^{ij}\dot{\g}_{ij}+\p_\vf\dot{\vf}+\p_\c\dot{\c}\right)\NO\\
&=&\frac{\pa\cs_{r_o}}{\pa r_o}+H+L,
\eea
where $\pa/\pa r_o$ denotes the {\em partial} derivative with respect to $r_o$. However, since $\cs_{r_o}$ is the on-shell
action, we must have $\dot{\cs}_{r_o}=L$ and so we conclude that
\be\label{H-J}
\frac{\pa\cs_r}{\pa r}+H=0.
\ee
For a generally covariant theory, like supergravity, the Hamiltonian vanishes identically since it is proportional to the constraints (\ref{constraints}). It follows that, as a consequence of the general covariance of the theory, the on-shell action does not depend explicitly on the radial coordinate, $r$, but only through the induced fields on the hypersurface $\S_r$. Moreover, we see that the Hamilton-Jacobi equation in a generally covariant theory is equivalent to the vanishing of the constraints, i.e.
\bea
\label{HJ-equations}
&&2\k^2\g^{-\frac12}\left(\left(\g_{ik}\g_{jl}-\frac{1}{d-1}\g_{ij}\g_{kl}\right)\frac{\d\cs_r}{\d\g_{ij}}\frac{\d\cs_r}{\d\g_{kl}}+\frac14\left(\frac{\d\cs_r}{\d\vf}\right)^2+\frac14Z^{-1}(\vf)\left(\frac{\d\cs_r}{\d\c}\right)^2\right)\NO\\
&&=-\frac{1}{2\k^2}\sqrt{\g}\left(R[\g]-\pa_i\vf\pa^i\vf-Z(\vf)\pa_i\chi\pa^i\chi+V(\vf)\right),\NO\\
&&-2D_j\left(\frac{\d\cs_r}{\d\g_{ij}}\right)+\left(\frac{\d\cs_r}{\d\vf}\right)\pa^i\vf+\left(\frac{\d\cs_r}{\d\c}\right)\pa^i\chi=0.
\eea
The task of the next subsection will be to systematically solve these equations is a certain asymptotic sense.

\subsection{Recursive solution of the Hamilton-Jacobi equation}

We now turn to the task of solving the Hamilton-Jacobi equations (\ref{HJ-equations}) in order to determine
the boundary term $S_b$ that makes the variational problem for the action (\ref{action-scalars}) at $r_o=\infty$
well defined. As was argued in \cite{Papadimitriou:2010as} and reiterated above, provided the variational 
problem is formulated within a well defined space of asymptotic solutions, the leading form of the boundary term, obtained as 
a solution of the Hamilton-Jacobi equation, will contain no transverse derivatives. It follows that the full 
solution of the Hamilton-Jacobi equation that corresponds to the boundary term we seek to determine admits an 
expansion in transverse derivatives. One can, therefore, try to solve the Hamilton-Jacobi equation by writing down an
ansatz containing all possible terms allowed by general covariance at each order in derivatives. Although this approach,
which was used in, for example, \cite{de_Boer:1999xf} in the case of asymptotically AdS gravity, generically suffices 
for simple cases in low spacetime dimension and for a limited number of fields, it quickly becomes prohibitively inefficient 
and cumbersome. In particular, this approach not only unnecessarily includes terms in the ansatz that may happen
to be absent in the particular theory, but also the number of equations one obtains for the undetermined functions in
the ansatz is greater than the number of functions to be determined, and so many equations are redundant. Instead, the 
approach we will develop here is a systematic recursive procedure for solving the Hamilton-Jacobi
equation, closer in spirit to the recursive method developed in \cite{Parry:1993mw}. In particular, the algorithm 
we will present computes systematically the $n$th term in the derivative expansion from lower terms, thus producing only the
terms that do appear in the actual solution of the Hamilton-Jacobi equation.

The basis of our algorithm is the following observation. Let us begin by writing Hamilton's principal function, $\cs_r$, as
\be
\cs_r=\int_{\S_r} d^dx \cl(\g,\vf,\c).
\ee
From the relations (\ref{HJ-momenta}) then we have
\be\label{variational-id}
\p^{ij}\d\g_{ij}+\p_\vf\d\vf+\p_\c\d\c=\d\cl+\pa_iv^i(\d\g,\d\vf,\d\c),
\ee
for some vector field $v^i(\d\g,\d\vf,\d\c)$. Since by construction the solution $\cs_r$ we are seeking admits a derivative
expansion as $r\to\infty$, $\cs_r$ increasingly approaches a solution of the form
\be\label{zero-order-sol}
\cs\sub{0}=\frac{1}{\k^2}\int_{\S_r}d^dx \sqrt{\g}U(\vf,\chi),
\ee
for some function $U(\vf,\chi)$. Given the zero order solution (\ref{zero-order-sol}), we can compute corrections to this action in a systematic
expansion in eigenfunctions of the operator\footnote{In the case of asymptotically AdS or dS gravity coupled to matter, it 
was proposed in \cite{Papadimitriou:2004ap} that one expands $\cs_r$ and the corresponding canonical momenta in eigenfunctions
of the dilatation operator, $\d_D$, obtained from the relation 
$\pa_r=\int d^dx\left(\dot\g_{ij}\frac{\d}{\d\g_{ij}}+\S_f\dot f\frac{\d}{\d f}\right)\xrightarrow{r\to\infty} \frac1\ell\d_D$, 
where $f$ denotes generic matter fields. In that case expanding in eigenfunctions of the dilatation operator is the most efficient 
recursive method to solve the Hamilton-Jacobi equation, because it amounts to solving the zero order problem (\ref{zero-order-eq})
and the one determining the higher derivative terms simultaneously. However, in more general cases the operator corresponding to 
the leading asymptotic behavior of $\pa_r$ is not very useful in practice since its eigenfunctions are not simple functions
of curvature invariants and matter fields. In most cases it is easier in practice to first solve the zero order problem (\ref{zero-order-eq}) to determine $U(\vf,\c)$ and 
then to expand in eigenfunctions of $\d_\g$. Of course in the case of asymptotically AdS or dS gravity the two approaches 
produce identical results, even though the two expansions differ order by order, the difference being that the latter expansion 
resums all zero derivative terms into the leading term.}
\be\label{delta-operator}
\d_\g=\int d^dx 2\g_{ij}\frac{\d}{\d\g_{ij}},
\ee
namely (dropping the subscript $r$ from now on) 
\be
\cs=\cs\sub{0}+\cs\sub{2}+\cs\sub{4}+\cdots,
\ee
where $\d_\g \cs\sub{2n}=(d-2n)\cs\sub{2n}$. It is easy to see that the resulting expansion is a derivative expansion.
Note that the operator (\ref{delta-operator}) agrees with the dilatation operator introduced in \cite{Papadimitriou:2004ap}
in the case of a constant potential, i.e. for the usual dilaton-axion system in conformal theories. However, the would-be
dilatation operator for an arbitrary dilaton potential would lead to an operator whose eigenfunctions are highly non-trivial
and so, from a practical point of view, would not serve as a good basis for expanding Hamilton's principal function. Moreover,
an expansion in eigenfunctions of the operator (\ref{delta-operator}) has the advantage of leading to {\em algebraic} in
the induced metric $\g_{ij}$ equations for determining the terms $\cs\sub{2n}$, which is the fundamental ingredient in our
algorithm.

Let us see how this works. Applying the general identity (\ref{variational-id}) to the
variation $\d_\g$ we obtain
\be
2\p\sub{2n}=(d-2n)\cl\sub{2n}+\pa_i v^i\sub{2n}.
\ee
Since $\cl$ is defined up to a total derivative, we can absorb the last term in this identity into $\cl\sub{2n}$
such that
\be\label{identity}
2\p\sub{2n}=(d-2n)\cl\sub{2n}.
\ee
The significance of this relation will become clear shortly, when we write down the equation determining $\cl\sub{2n}$.
Since (\ref{identity}) holds only for a certain choice of total derivative terms in $\cl\sub{2n}$, in principle we
should keep track of total derivative terms as well. However, as we shall see, this will not be necessary in our
recursive procedure. In particular, we will determine $\cl\sub{2n}$ at each order up to total derivative terms. The canonical
momenta at this order can then be obtained by differentiating $\cs\sub{2n}=\int d^dx \cl\sub{2n}$. Since total derivative
terms do not influence the canonical momenta, we will still get the correct expressions for the canonical momenta.

Inserting the leading term (\ref{zero-order-sol}) of the above expansion in to the Hamilton-Jacobi equation
(\ref{HJ-equations}) we find that the function $U(\vf,\c)$ satisfies the equation
\be
\label{zero-order-eq}
(\pa_\vf U)^2+Z^{-1}(\vf)(\pa_\chi U)^2-\frac{d}{d-1}U^2+V(\vf)=0.
\ee
However, using the leading term (\ref{zero-order-sol}) in the relations (\ref{HJ-momenta}) and identifying the
momenta from the expressions (\ref{momenta-scalars}) we find that to leading order the induced metric takes the
form\footnote{We gauge-fix the lapse and shift functions to $N=1$ and $N_i=0$.}
\be\label{leading-metric}
\g_{ij}=e^{2A}\bar g\sub{0}_{ij}(x),
\ee
where $\bar g\sub{0}_{ij}(x)$ is an arbitrary metric on the boundary and the fields $A$, $\vf$ and $\c$ satisfy
the flow equations (\ref{flow_eqs}), but with $W$ replaced with $U$. But then, these first order equations in
combination with (\ref{zero-order-eq}) imply the second order equations (\ref{dw_eqs}). As we have seen above,
the second order equation for the axion is integrable, giving in this case
\be
\frac{\pa U(\vf,\c)}{\pa\c}=\tilde{c}e^{-dA},
\ee
which is the analogue of (\ref{integrated-axion}). We therefore see that any $\c$ dependence in $U$ leads to a finite
contribution in Hamilton's principal function $\cs$ and hence we can set the integration constant $\tilde{c}=0$. We
conclude that the function $U(\vf)$ required to make the variational problem well defined can be taken, without loss of
generality, to be {\em independent of the axion} $\c$. As we will see shortly, this leads to a significant simplification of
the analysis to determine the required higher derivative terms in $\cs$. 

In order to determine the leading solution (\ref{zero-order-sol}) of the Hamilton-Jacobi equation, therefore, one needs to solve the 
reduced equation 
\be
\label{zero-order-eq-reduced}
(\pa_\vf U)^2-\frac{d}{d-1}U^2+V(\vf)=0.
\ee
For an arbitrary potential this equation can be transformed into an Abel's equation of the first kind \cite{Papadimitriou:2004ap},
which is generically non-integrable. However, we need not find the general solution of this equation. Any solution 
of this equation that ensures that $\vf$ has the desired general asymptotics via the relation 
\be
\dot\vf=\pm\sqrt{\frac{d}{d-1}U^2(\vf)-V(\vf)},
\ee 
suffices. In particular, if $U$ is such a solution and it is not isolated in the space of solutions\footnote{See Appendix
\ref{constant-potential} for an example where this happens.}, then it is easy to see that any solution in the {\em vicinity} 
of $U$ is of the form $U+\e \D U$, where $\e$ is an infinitesimal parameter and $\D U=\co(\exp(-dA))$. Hence, the difference 
between two such solutions only contributes a finite term in $\cs\sub{0}$, and it is therefore irrelevant. The reason for 
restricting this argument to the vicinity of the original solution $U$ is that there exist solutions at infinite parametric 
distance from $U$ that change the leading asymptotic behavior of $\vf$. Such solutions are excluded by the requirement that 
$\vf$ has the correct asymptotics.       

Now that we have determined that $U(\vf)$ is independent of the axion, inserting the above expansion of Hamilton's principal function in the Hamiltonian constraint and matching terms of equal $\d_\g$ eigenvalue we obtain for the higher
order terms
\be
\label{linear-eq}
U'(\vf)\frac{\d}{\d\vf}\int d^dx\cl\sub{2n}-\left(\frac{d-2n}{d-1}\right)U(\vf)\cl\sub{2n}=\car\sub{2n},\quad n>0,
\ee
where
\bea\label{sources}
\car\sub{2}&=&-\frac{1}{2\k^2}\sqrt{\g}\left(R[\g]-\pa_i\vf\pa^i\vf-Z(\vf)\pa_i\chi\pa^i\chi\right),\NO\\
\car\sub{2n}&=&-2\k^2\g^{-\frac12}\sum_{m=1}^{n-1}\left(\p\sub{2m}^i_j\p\sub{2(n-m)}^j_i-\frac{1}{d-1}\p\sub{2m}\p\sub{2(n-m)}
\right.\NO\\
&&\left.+\frac14 \p_\vf\sub{2m}\p_\vf\sub{2(n-m)}+\frac14 Z^{-1}(\vf)\p_\chi\sub{2m}\p_\chi\sub{2(n-m)}\right), \quad n>1.
\eea
Importantly, these are {\em linear} equations and only involve a derivative w.r.t. the dilaton, $\vf$, and {\em not} the induced metric $\g_{ij}$ or the axion. The absence of a derivative w.r.t. the induced metric is due to the relation (\ref{identity}), while
the absence of a derivative w.r.t. the axion is because we have shown that $U$ only depends on the dilaton. We will now solve
these equations up to the order required in order to determine the boundary term $S_b$ for $d=4$. In the analysis
below we will take $U'(\vf)\neq 0$, which is guaranteed by the requirement of a logarithmic running for the dilaton. However,
the case of a constant potential $V$ and, consequently, constant $U$ arises in conformal theories such as $\cn=4$ super Yang-Mills
and so it is interesting on its own right. For completeness we discuss this case in Appendix \ref{constant-potential}.

Turning now to the case $U'(\vf)\neq 0$, we notice that the linear equation (\ref{linear-eq}) for $\cl\sub{2n}$, $n>0$, admits the homogeneous solution
\be
\cl^{hom}_{(2n)}=\cf\sub{2n}[\g,\c]\exp\left(\left(\frac{d-2n}{d-1}\right)\int^\vf
\frac{d\bar{\vf}}{U'(\bar{\vf})}U(\bar{\vf})\right),
\ee
where $\cf\sub{2n}[\g,\c]$ is a covariant function of the induced metric and the axion of weight $d-2n$. However, this solution contributes only to finite local counterterms and can be ignored. To see this notice that the leading order solution (\ref{zero-order-sol}) implies via the Hamilton-Jacobi relations (\ref{HJ-momenta}) that
the induced metric takes to leading order the form (\ref{leading-metric}) with
\be\label{warp-factor}
A=-\frac{1}{d-1}\int^\vf\frac{d\bar{\vf}}{U'(\bar{\vf})}U(\bar{\vf}).
\ee
Now, by construction, $\cf\sub{2n}[\g,\c]$ has weight $d-2n$ and so
\be
\cl^{hom}_{(2n)}\sim e^{(d-2n)A}\times e^{-(d-2n)A} = finite.
\ee
We are, therefore, only interested in the {\em inhomogeneous} solution of (\ref{linear-eq}). In other words, formally,
\be\label{implicit-solution}
\cl\sub{2n}=e^{-(d-2n)A(\vf)}\int^\vf
\frac{d\bar{\vf}}{U'(\bar{\vf})}e^{(d-2n)A(\bar{\vf})}\car\sub{2n}(\bar{\vf}).
\ee
This integral is well defined if $\car\sub{2n}$ does not involve derivatives of the dilaton, $\vf$, but
it requires some caution when it does. In general, we can write more precisely
\be\label{ln}
\framebox[2.0in]{\rule[-.1in]{0in}{0in}
\begin{minipage}{3.8in}
\begin{equation*}
\cl\sub{2n}=e^{-(d-2n)A(\vf)}F\sub{2n},
\end{equation*}
\end{minipage}
}
\ee
where $F\sub{2n}$ satisfies
\be\label{integration-formula}
\framebox[4.5in]{\rule[-.1in]{0in}{0in}
\begin{minipage}{5.8in}
\begin{equation*}
\frac{\d\vf}{U'(\vf)}e^{(d-2n)A(\vf)}\car\sub{2n}(\vf)=\d_\vf F\sub{2n}+e^{(d-2n)A(\vf)}\pa_i v\sub{2n}^i(\vf,\d\vf),
\end{equation*}
\end{minipage}
}
\ee
for some vector field $v\sub{2n}^i(\vf,\d\vf)$. In Table \ref{f-integration} we have listed the local functional
$F\sub{2n}(\vf)$ for a number of generic source terms $\car\sub{2n}(\vf)$ that we will need for our calculation.
A detailed derivation of the formulas given in Table \ref{f-integration} is provided for the reader's convenience in
Appendix \ref{functional-integration}. Both in Table \ref{f-integration} and in Appendix \ref{functional-integration}
we make extensive use of the short-hand notation
\be
\fint_{n,m}^\vf\equiv \left(A'\right)^m e^{-(d-2n)A}\int^\vf\frac{d\bar\vf}{U'}e^{(d-2n)A}\left(A'\right)^{-m},
\ee
where $A(\vf)$ is given by (\ref{warp-factor}).
\begin{table}
\[\begin{array}{|l|l|}
\hline\hline
\car\sub{2n} & e^{-(d-2n)A}F\sub{2n}\\ \hline &\\

r_{1^m}(\vf)t^{i_1i_2\ldots i_{m}}\pa_{i_1}\vf\pa_{i_2}\vf\ldots\pa_{i_m}\vf &
\fint_{n,m}^\vf r_{1^m}(\bar\vf)t^{i_1i_2\ldots i_{m}}\pa_{i_1}\vf\pa_{i_2}\vf\ldots\pa_{i_m}\vf\\ &\\

r_2(\vf)t^{ij}D_iD_j\vf & \fint_{n,1}^\vf r_{2}(\bar\vf)t^{ij}D_iD_j\vf\\ &
-\fint_{n,2}^\vf U'A'\pa_{\bar\vf}^2\left(\frac{1}{A'}\right)\fint_{n,1}^{\bar\vf}r_{2}(\tilde\vf)t^{ij}\pa_i\vf\pa_j\vf\\ &\\

\left(r_{1^22}(\vf)t_1^{ijkl}+s_{1^22}(\vf)t_2^{ijkl}\right)\pa_i\vf\pa_j\vf D_kD_l\vf &
\fint_{n,3}^\vf s_{1^22}(\bar\vf)t_2^{ijkl}\pa_i\vf\pa_j\vf D_kD_l\vf \\ & \\

\left(r_{2^2}(\vf)t_1^{ijkl}+s_{2^2}(\vf)t_2^{ijkl}\right)D_iD_j\vf D_kD_l\vf &
\left(\fint_{n,2}^\vf r_{2^2}(\bar\vf)t_1^{ijkl}+\fint_{n,2}^\vf s_{2^2}(\bar\vf)t_2^{ijkl}\right)D_iD_j\vf D_kD_l\vf \\&
-2\fint_{n,3}^\vf U'A'\pa_{\bar\vf}^2\left(\frac{1}{A'}\right)
\fint_{n,2}^{\bar\vf} s_{2^2}(\tilde\vf)t_2^{ijkl}\pa_i\vf\pa_j\vf D_kD_l\vf \\

&\\ \hline\hline
\end{array}\]
\caption{The result of the functional integration described by (\ref{integration-formula}) for various source terms. 
Here, $t^{i_1i_2\ldots i_{m}}$ and $t^{ij}$ are arbitrary totally symmetric tensors independent of $\vf$, while 
$t_1^{ijkl}=\frac13\left(\g^{ik}\g^{jl}+\g^{il}\g^{jk}+\g^{ij}\g^{kl}\right)$,   
$t_2^{ijkl}=\frac13\left(\g^{ik}\g^{jl}+\g^{il}\g^{jk}-2\g^{ij}\g^{kl}\right)$. Moreover, $F_{(2n)}$ is given up to terms of 
the form $e^{(d-2n)A}D_i\cu^i$, for some vector field $\cu^i$. Since $\cl_{(2n)}$ is related to $F_{(2n)}$ as in
(\ref{ln}), $\cu^i$ corresponds to a total derivative term in $\cl_{(2n)}$ and so we need not determine this term explicitly. 
Note that the source term $r_{1^22}$ only contributes to $\cu^i$ and so it can be ignored. A detailed derivation of these results
is given in Appendix \ref{functional-integration}.}
\label{f-integration}
\end{table}

\begin{figure}
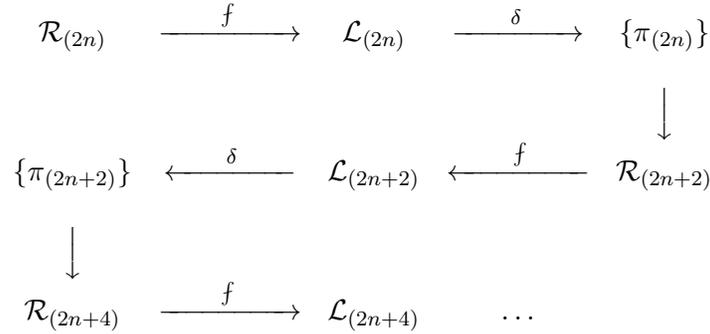

\[\begin{array}{ccccc}
\car\sub{2n} & \xrightarrow{\phantom{more}\fint\phantom{more}} & \cl\sub{2n} & \xrightarrow{\phantom{more}\d\phantom{more}} & \{\p\sub{2n}\} \\
&&&&\\
&&&&\Big\downarrow \\
\{\p\sub{2n+2}\} & \xleftarrow{\phantom{more}\d\phantom{more}} & \cl\sub{2n+2} & \xleftarrow{\phantom{more}\fint\phantom{more}} & \car\sub{2n+2} \\&&&&\\
\Big\downarrow&&&& \\
\car\sub{2n+4} &\xrightarrow{\phantom{more}\fint\phantom{more}} & \cl\sub{2n+4} &\ldots &
\end{array}\]
\caption{A schematic illustration of the algorithm used to determine Hamilton's principal function iteratively. Here,
the operators $\fint$ and $\d$ stand respectively for the functional integration defined by the formula in
(\ref{integration-formula}) and for functional differentiation with respect to the induced fields. }
\label{algorithm}
\end{figure}

The integration formula (\ref{integration-formula}) allows us to develop an algorithmic procedure for evaluating Hamilton's
principal function, $\cs$, iteratively. Namely, given the source term of equation (\ref{linear-eq}) at order $n$,
(\ref{integration-formula}) is used to obtain $\cl\sub{2n}$ up to an irrelevant total derivative term. Differentiating
this with respect to the various induced fields gives the corresponding momenta at that order. Finally, using these momenta,
as well as those of lower orders, one determines the source term at order $n+1$ via (\ref{sources}). The procedure is
then repeated to the desired order. This algorithm is schematically outlined in Fig. \ref{algorithm}.
Let us now carry out this general algorithm to order $n=2$, which is sufficient for evaluating the boundary term
for the action (\ref{action-scalars}) in $d=4$. In general, at each order $n$ the source term, $\car\sub{2n}$ and
the corresponding inhomogeneous solution, $\cl\sub{2n}$, of the functional equation (\ref{linear-eq})
can be written respectively in the form
\be\label{source-parameterization}
\car\sub{2n}=-\frac{1}{2\k^2}\sqrt{\g}\sum_{I=1}^{N_n}c^I_n(\vf) \ct^I_n,
\ee
and
\be\label{solution-parameterization}
\cl\sub{2n}=-\frac{1}{2\k^2}\sqrt{\g}\sum_{I=1}^{N_n}\cp^I_n(\vf) \ct^I_n,
\ee
where $c^I_n(\vf)$ and $\cp^I_n(\vf)$ are scalar functions of $\vf$ and $\ct^I_n$ are quantities involving
fields other than $\vf$, as well as derivatives of $\vf$. The number $N_n$  is the number of such quantities at each order $n$.
For $n=1$ the source $\car\sub{2}$ is given in (\ref{sources}). Applying the results in Table \ref{f-integration}
to this sources we obtain $\cl\sub{2}$. The result is summarized in Table \ref{order-1}.

\begin{table}
\[\begin{array}{|c|l|l|l|}
\hline\hline
I &  \ct^I_1& c^I_1(\vf) &\cp_1^I(\vf)\\ \hline & & & \\

1 & R & 1 & \fint_{1,0}^\vf 1\equiv-2\X(\vf)\\&&&\\

2 & \pa_i\vf\pa^i\vf& -1 & \fint_{1,2}^\vf (-1)\equiv-M(\vf)\\ &&&\\

3 & \pa_i\c\pa^i\c & -Z(\vf) & \fint_{1,0}^\vf (-Z(\bar\vf))\equiv-\Th(\vf)\\&&&\\

\hline\hline
\end{array}\]
\caption{Summary of the source terms (\ref{source-parameterization}) and the corresponding solution (\ref{solution-parameterization}) 
of the Hamilton-Jacobi equation at order $n=1$. The source term is determined iteratively 
as in (\ref{sources}), while $\cl_{(2n)}$ are determined by solving the linear equations 
(\ref{linear-eq}) via the integration procedure described in the text.}
\label{order-1}
\end{table}

In order to move to the next order we first need to evaluate the canonical momenta at order $n=1$ by
differentiating $\cl\sub{2}$ with respect ot the induced fields. After a little algebra we obtain:
\bea
\p\sub{2}^{ij}&=&-\frac{1}{\k^2}\sqrt{\g}\left(\X R^{ij}-\X'D^iD^j\vf+\frac12 (M-2\X'')\pa^i\vf\pa^j\vf
+\frac12\Th\pa^i\chi\pa^j\chi\right.\NO\\
&&\left.-\frac12\g^{ij}\left(\X R-2\X'\square_\g\vf+\frac12 (M-4\X'') \pa_k\vf\pa^k\vf+\frac12\Th\pa_k\chi\pa^k\chi\right)\right), \NO\\
\p_\vf\sub{2}&=&-\frac{1}{\k^2}\sqrt{\g}\left(\frac12M'\pa_i\vf\pa^i\vf+M\square_\g\vf-\frac12\Th'\pa_i\c\pa^i\c-\X'R\right),\NO\\
\p_\c\sub{2}&=&-\frac{1}{\k^2}\sqrt{\g}\left(\Th\square_\g\c+\Th'\pa_i\vf\pa^i\c\right).
\eea
Inserting these in the expression for $\car\sub{4}$ in (\ref{sources}) results in an explicit expression for
the source term at order $n=2$. This is a rather complicated source, involving 20 different terms, and can be read out from
the first three columns of Table \ref{order-2}. Using the results of Table \ref{f-integration} we then obtain
the fourth column in Table \ref{order-2}, giving $\cl\sub{4}$. This table is the main result of this paper providing
together with Table \ref{order-1}, 
the boundary term necessary to make the variational problem of a generic action of the form (\ref{action-scalars}) well defined
in any dimension up to and including $d+1=5$. 

Before we apply this general result to IHQCD, a technical comment is due. The boundary terms listed in Tables \ref{order-1} and
\ref{order-2} generically have poles at $d=2$ or $d=4$. This happens whenever there is a conformal anomaly, or, in terms
of the bulk language, whenever the boundary term required to make the variational problem well defined breaks the bulk 
diffeomorphisms corresponding to translations in the radial coordinate. The way to handle these poles is to relate 
the radial cut-off, $r_0$, to the parameter $d$, as\footnote{The radial cut-off $r_0$ and the boundary dimension
$d$ are arbitrary parameters in the radial Hamiltonian formalism. However, the fact that this formalism, without relying on 
any additional input from the asymptotic expansions, can handle conformal anomalies in $d_*$ dimensions by relating the radial
cut-off to the parameter $d$ suggests that a radial cut-off regularization corresponds to 
dimensional regularization in the dual field theory. Taken at face value, this identification has profound consequences for 
the physical significance of the holographic direction.} $r_0=1/(d-d_*)$, where $d_*$ is the dimension of the boundary. 
An example of this issue arises in the case of a constant dilaton potential, discussed in Appendix \ref{constant-potential}.  
\begin{landscape}
\begin{table}
\[\begin{array}{|c|l|l|l|}
\hline\hline
I &  \ct^I_2& c^I_2(\vf) &\cp^I_2(\vf) \\
\hline & & &\\
1 & R^{ij}R_{ij} & 4\X^2 & \fint_{2,0}^\vf 4\X^2\\&&&\\

2 & R^2& \X'^2-\frac{d}{d-1}\X^2 &\fint_{2,0}^\vf(\X'^2-\frac{d}{d-1}\X^2)\\ &&&\\

3 & D^iD^j\vf D_iD_j\vf & 4\X'^2 & 4\fint_{2,2}^\vf\X'^2\\&&&\\

4 & D^iD^j\vf\pa_i\vf\pa_j\vf & -4\X'\left(M-2\X''\right) & \frac23\fint_{2,3}^\vf ( 4\X'(3\X''-M)-MM'
-2U'A'\pa_{\bar\vf}^2\left(\frac{1}{A'}\right)\fint_{2,2}^{\bar\vf}(6\X'^2-M^2))\\&&&\\

5 & \square_\g\vf\pa_i\vf\pa^i\vf & MM'+2\X'\left(M-4\X''\right) & -\frac23\fint_{2,3}^\vf ( 4\X'(3\X''-M)-MM'
-2U'A'\pa_{\bar\vf}^2\left(\frac{1}{A'}\right)\fint_{2,2}^{\bar\vf}(6\X'^2-M^2))\\&&&\\

6 & \left(\square_\g\vf\right)^2 & M^2-4\X'^2 & \fint_{2,2}^\vf \left(M^2-4\X'^2\right)\\&&&\\

7 & \left(\pa_i\vf\pa^i\vf\right)^2 & \frac14M'^2+\frac{(3d-4)}{4(d-1)}M^2-2M\X'' & \fint_{2,4}^\vf (\frac14M'^2+\frac{(3d-4)}{4(d-1)}M^2-2M\X'')\\&&&\\

8 & \left(\pa_i\c\pa^i\c\right)^2 & \frac14\Th'^2+\frac{(3d-4)}{4(d-1)}\Th^2 & \frac14\fint_{2,0}^\vf(\Th'^2+\frac{(3d-4)}{(d-1)}\Th^2)\\&&&\\

9 & R^{ij}D_iD_j\vf & -8\X\X' & -8\fint_{2,1}^\vf \X\X'\\&&&\\

10 & R^{ij}\pa_i\vf\pa_j\vf & 4\X\left(M-2\X''\right) & 4\fint_{2,2}^\vf \left(\X(M-2\X'')+2U'A'\pa_{\bar\vf}^2\left(\frac{1}{A'}\right)\fint_{2,1}^{\bar\vf}\X\X'\right)\\&&&\\

 \hline\hline
\end{array}\]
\end{table}

\begin{table}
\[\begin{array}{|c|l|l|l|}
\hline
\hline & & &\\

11 & R\square_\g\vf & -2\X'\left(M-2\X\right) &-2\fint_{2,1}^\vf \X'\left(M-2\X\right)\\&&&\\

12 & R\pa_i\vf\pa^i\vf & -\X'M'-\frac{d}{d-1}\X M+4\X\X'' &\fint_{2,2}^\vf (\rule{0cm}{.4cm}-\X'M'-\frac{d}{d-1}\X M+4\X\X''
+2U'A'\pa_{\bar\vf}^2\left(\frac{1}{A'}\right)\fint_{2,1}^{\bar\vf}\X'(M-2\X))\\&&&\\

13 & R^{ij}\pa_i\c\pa_j\c & 4\X\Th &\fint_{2,0}^\vf 4\X\Th\\&&&\\

14 & D^iD^j\vf\pa_i\c\pa_j\c & -4\X'\Th & -4\fint_{2,1}^\vf \X'\Th\\&&&\\

15 & \left(\pa_i\vf\pa^i\c\right)^2 & Z^{-1}\Th'^2+2\left(M-2\X''\right)\Th & \fint_{2,2}^\vf (Z^{-1}\Th'^2+2\left(M-2\X''\right)\Th+4U'A'\pa_{\bar\vf}^2\left(\frac{1}{A'}\right)\fint_{2,1}^{\bar\vf}\X'\Th)\\&&&\\

16 & \square_\g\vf\pa_i\c\pa^i\c & 2\X'\Th-M\Th' &\fint_{2,1}^\vf \left(2\X'\Th-M\Th'\right)\\&&&\\

17 & R\pa_i\c\pa^i\c & \X'\Th'-\frac{d}{d-1}\X \Th & \fint_{2,0}^\vf (\X'\Th'-\frac{d}{d-1}\X \Th)\\&&&\\

18 & \pa_i\vf\pa^i\vf\pa_j\c\pa^j\c & \frac12\Th\left(4\X''-\frac{d}{d-1}M\right)-\frac12M'\Th' & \frac12\fint_{2,2}^\vf (\Th(4\X''-\frac{d}{d-1}M)
-M'\Th'-2A'U'\pa_{\bar\vf}^2\left(\frac{1}{A'}\right)\fint_{2,1}^{\bar\vf}(2\X'\Th-M\Th'))\\&&&\\

19 & \left(\square_\g\c\right)^2 & Z^{-1}\Th^2 &\fint_{2,0}^\vf Z^{-1}\Th^2\\&&&\\

20 & \square_\g\c \pa_i\vf\pa^i\c & 2Z^{-1}\Th\Th' & \fint_{2,1}^\vf 2Z^{-1}\Th\Th'\\
 &&&\\ \hline\hline
\end{array}\]
\caption{This table summarizes the source terms (\ref{source-parameterization}) and the corresponding solution (\ref{solution-parameterization}) of 
the Hamilton-Jacobi equation at order $n=2$. The 
solution $\cl_{(4)}$ of the Hamilton-Jacobi equation at order $n=2$ can simply be read off the fourth column via the relation
(\ref{solution-parameterization}). This table is the main result of this paper providing, together with Table \ref{order-1}, 
the boundary term necessary to make the variational problem of a generic action of the form (\ref{action-scalars}) well defined
in any dimension up to and including $d+1=5$.}
\label{order-2}
\end{table}
\end{landscape}

\section{Application to IHQCD}
\label{IHQCD-application}
\setcounter{equation}{0}

The results presented in the previous section are applicable to any action of the form (\ref{action-scalars}). In particular,
Tables \ref{order-1} and \ref{order-2} provide the general boundary term that renders the variational problem of any action
of the form (\ref{action-scalars}) well posed for boundary dimension up to and including $d=4$. However, additional assumptions 
that enter into specific models often lead to significant simplification of this boundary term. In this section we will apply 
these general results to the specific problem of IHQCD \cite{Gursoy:2007cb} described in the Introduction. The 
significance of this boundary term will be demonstrated by deriving asymptotic expansions analogous to the 
Fefferman-Graham expansion for asymptotically AdS gravity \cite{FG}, providing general expressions for the one-point functions 
of the dual operators in terms of the coefficients of these expansions, and finally giving the correct holographic Ward identities 
for IHQCD.    

In deriving the boundary term in Tables \ref{order-1} and \ref{order-2} we have assumed that the leading asymptotics of the 
induced fields follows from a leading solution (\ref{zero-order-sol}) of the Hamilton-Jacobi equation which does not 
involve transverse derivatives. However, no specific asymptotics was assumed. Taking into account the 
leading asymptotic form of the induced fields of the IHQCD model allows for various simplifications of the result of the previous 
section. To see what simplifications occur let us start by writing $\vf=\x \log\l$, for some non-zero constant $\x$. Then, by abuse of notation, 
we will correspondingly write $U(\l)$, $A(\l)$, etc. as functions of $\l$. As we have argued above, $U$ is a function of the dilaton 
$\l$ only and not of the axion. Moreover, it is a solution of equation (\ref{zero-order-eq}), which now becomes
\be\label{zero-order-eq-lambda}
\frac{d}{d-1}U^2-\x^{-2}\l^2\left(\frac{\pa U}{\pa\l}\right)^2=V(\l).
\ee

At this point we need to invoke some information about the asymptotics. First, recall that we want to consider potentials of the form
(\ref{generic-potential}) as $\l\to0$. Moreover, $U(\l)$ defines the leading form of the metric and dilaton asymptotics
via the flow equations (these are the same as (\ref{flow_eqs}) but with $W$ replaced by $U$, and follow from
combining the relations (\ref{momenta-scalars}) and (\ref{HJ-momenta}) for the canonical momenta)
\be
\dot A=-\frac{1}{d-1}U(\l),\quad \dot{\l}=\x^{-2}\l^2\frac{\pa U(\l)}{\pa\l}.
\ee
The leading metric and dilaton asymptotics adopted in the model of IHQCD is \cite{Gursoy:2007cb}
\be\label{IHQCD-asymptotics}
\dot A\sim \ell^{-1},\quad \dot\l\sim -b_0\ell^{-1}\l^2+b_1\ell^{-1}\l^3,\,\, {\rm as}\,\, \l\to 0
\ee
where the constants $b_0$ and $b_1$ were argued in \cite{Gursoy:2007cb} to be related respectively to the one and two-loop 
perturbative beta function coefficients. According to that argument, subleading corrections to the asymptotic form of the dilaton 
correspond to higher loop coefficients in the perturbative beta function, which are scheme dependent. In that sense,
the terms in (\ref{IHQCD-asymptotics}) are universal in the model of IHQCD, and any dilaton potential $V(\l)$ chosen must be such
that it is compatible with this leading asymptotics in the UV, while subleading terms in the UV expansion of the potential
can differ for different choices of the dilaton potential. 

One should keep in mind, however, that the beta function of any operator whose coupling transforms inhomogeneously 
under scale transformations is scheme dependent and, therefore, not a physical observable. In other words, the 
beta function of any operator that acquires anomalous dimension is scheme dependent. For such operators the 
scaling dimension $\g_\co$ is not constant along the RG flow. Only the value of this scaling dimension at 
the fixed points is a physical observable. Nevertheless, the combination $\beta_\mathcal{O}\cdot\mathcal{O}(x)$ {\em is} a physical 
observable\footnote{We are grateful to Elias Kiritsis for pointing this out to us.}, since this combination 
appears in the trace of the stress tensor, whose scaling dimension does not renormalize. As we shall see below from the 
trace Ward identity, the bulk dilaton $\l$ should be thought of as the holographic dual to the operator $\b(\l_{YM})\Tr F^2$. 
Isolating the beta function factor in this operator is a scheme-dependent procedure, but the product transforms 
homogeneously under the renormalization group flow. It is precisely this combination that appears in the trace Ward identity.  
In fact, this is a generic mechanism of how one can accommodate operators with running scaling dimensions in 
a supergravity setting, thus going beyond the realm of operators with protected dimensions.  
   
In order to ensure that the leading asymptotic behavior of the metric and dilaton is of the form (\ref{IHQCD-asymptotics})
we take the function $U(\l)$ to be of the form    
\be 
\label{U-UV}
U(\l)=-\frac{d-1}{\ell}-\frac{\x^2b_0}{\ell}\l+\frac1\ell U_2\l^2+\frac1\ell U_\a\l^\a+o(\l^\a),\quad \a>2,
\ee
where $b_1=2\x^{-2}U_2$, and we have introduced a term of order $\l^\a$, $\a>2$,  
in order to allow for potentials with non-integer powers of $\lambda$, as were considered e.g. in \cite{Gursoy:2010fj,Panero:2009tv}.
However, it should be emphasized that the {\em} full closed form of the function $U(\l)$ is required in order to 
construct the necessary boundary term that makes the variational problem well defined. In particular, to write 
down explicitly the boundary term one necessarily needs an exact solution of (\ref{zero-order-eq-lambda}). Such an
exact solution is of course dependent on the particular choice of dilaton potential. Since we want to keep 
the discussion general here, we will not fix the dilaton potential or the function $U(\l)$, beyond the requirement
that as $\l\to0$ it behaves as in (\ref{U-UV}). Once a dilaton potential is specified, one then will simply need
to find an exact solution of (\ref{zero-order-eq-lambda}) that behaves asymptotically as in (\ref{U-UV}) and use it 
in the expressions below, all of which are expressed in terms of a generic $U(\l)$. Of course, such a solution 
will not be unique since it can be shown \cite{Papadimitriou:2006dr} that if $U_0(\l)$ is such a solution, then there is a 
continuous family of deformations of this solution, with the deformation behaving as $\l^{d\n}\exp(-d/b_0\l)$ as $\l\to 0$,
where the value of the exponent $\n$ is given below. However, {\em any} member of this continuous family of solutions 
is equally good for the purposes of evaluating the boundary term since any difference arising from this deformation only 
contributes finite terms to the boundary term.

\subsection{Boundary term for IHQCD}

In this section we will make use of the asymptotic form (\ref{U-UV}) of the function $U(\l)$ in order to 
simplify the general boundary term derived in the previous section. To this end we start by noting that, depending
on the value of the exponent $\a$, the function $A(\l)$ defined in (\ref{warp-factor}) has the following asymptotic 
behavior as $\l\to 0$ 
\bea
e^{A(\l)}=\left\{\begin{matrix}
          \l^{-\n}e^{\frac{1}{b_0\l}}\left(1-\frac{\a U_\a}{(\a-2)\x^2b_0^2}\l^{\a-2}+o\left(\l^{\a-2}\right)\right), & \a<3,\\
          \l^{-\n}e^{\frac{1}{b_0\l}}\left(1-\frac{1}{b_0}\left(\left(\frac{2U_2}{\x^2b_0}\right)^2-\frac{U_2}{d-1}+\frac{3U_3}{\x^2b_0}\right)\l+o\left(\l\right)\right), & \a=3,\\    
	  \l^{-\n}e^{\frac{1}{b_0\l}}\left(1-\frac{1}{b_0}\left(\left(\frac{2U_2}{\x^2b_0}\right)^2-\frac{U_2}{d-1}\right)\l+o\left(\l\right)\right), & \a>3,
      \end{matrix}
\right.
\eea
where $\n=\frac{\x^2}{d-1}+\frac{2U_2}{\x^2b_0^2}$. Using these asymptotic expansions and the integral identity 
\bea
\int^\l d\l' \l'^{\m-2}e^{\frac{\om}{\l'}}
=\left\{\begin{matrix}
         -\frac{1}{\om}e^{\frac{\om}{\l}}\l^{\m}
\left(\sum_{n=0}^{n_o}\frac{\G(n+\m)}{\G(\m)}\left(\frac{1}{\om}\right)^n\l^n+\co\left(\l^{n_o+1}\right)\right),& \om>0,\; \forall n_o\in\mathbb{N},\\
\frac{1}{\m-1}\l^{\m-1}+const., & \om=0,\; \m\neq 1,\\
\log\l+const., & \om=0,\; \m=1,
        \end{matrix}\right.
\eea
we can determine the asymptotic form of all functions listed in Tables \ref{order-1} and \ref{order-2}. These functions involve integrals of the form
\be\label{generic-integral}
\fint_{n,m}^\l\l^\D \equiv \x^2\left(\l A'(\l)\right)^me^{-(d-2n) A(\l)}
\int^\l \frac{d\bar{\l}}{\bar{\l}^2 U'(\bar{\l})}e^{(d-2n) A(\bar{\l})}\left(\bar\l A'(\bar\l)\right)^{-m}\bar{\l}^\D,
\ee
whose asymptotic form we tabulate in Table \ref{integral-asymptotics}. 
\begin{table}
\[\begin{array}{|c|c|c|}
\hline\hline
d-2n & \D+m & \fint_{n,m}^\l\l^\Delta\\ \hline  && \\

  >0 & any & \frac{\ell \l^\D}{d-2n}\left(1+\left(\frac{\D+m}{d-2n}-\frac{\x^2}{d-1}\right)b_0\l+\co(\l^2)\right)\\&&\\

  0 & \neq0,1&-\frac{\ell\l^{\D-1}}{b_0(\D-1+m)}\left(1+\frac{1}{\D+m}\left((\D-1+2m)\frac{2U_2}{\x^2b_0}
+m\frac{\x^2b_0}{d-1}\right)\l+\co(\l^2)\right)\\ & &\\

0 & 1 & -\frac{\ell\l^{\D-1}}{b_0}\left(\log\l+\co(\l^0)\right)\\&&\\

0 & 0 & \frac{\ell\l^{\D-1}}{b_0}\left(1-\left(\frac{2U_2}{\x^2b_0}+\D b_0\n\right)\l\log\l+\co(\l)\right)\\&&\\

\hline\hline
\end{array}\]
\caption{The asymptotic form of the integral (\ref{generic-integral}) for a function $U(\l)$ that behaves 
as in (\ref{U-UV}) as $\l\to 0$. }
\label{integral-asymptotics}
\end{table}
With these results one can now determine the asymptotic form of the functions in Tables \ref{order-1} and \ref{order-2}, which we
present respectively in Tables \ref{IHQCD-order-1} and \ref{IHQCD-order-2}. In deriving these asymptotic forms we have made repeated
use of the crucial fact that $\pa_i\l=\co(\l^2)$. This follows from the general asymptotic form of the dilaton $\l$ and
will be derived in the next section.

Note that even though we have listed the leading asymptotic behavior of 
the order $n=1$ terms in Table \ref{IHQCD-order-1}, we actually need the exact closed form expressions of the functions 
$M(\l)$, $\X(\l)$ and $\Th(\l)$ in the boundary term. The reason is quite obvious. Namely, the expansion of the solution
of the Hamilton-Jacobi equation in eigenfunctions of the operator $\d_\g$, i.e. the derivative expansion, is
an expansion in powers of $e^{-2A}\sim\l^{2\n}\exp(-2/b_0\l)$. This expansion, therefore, is non-perturbative in $\l$. However,
at each order in the derivative expansion the boundary term contains an infinite expansion in powers of $\l$. In other words,
the expansion of the boundary term is a double expansion.\footnote{Generically, there are logarithmic in $\l$ terms as well, but
these can be included with the powers in $\l$ since they do not affect the counting of the derivative expansion.} 
Clearly, at 
orders $n=0$ and $n=1$ in the derivative expansion we must keep the {\em entire} perturbative expansions in $\l$, since there are 
divergences coming from any power of $\l$ no matter how large. This is the reason why the full closed form expressions for $U(\l)$ 
at order $n=0$ and of $M(\l)$, $\X(\l)$ and $\Th(\l)$ at order $n=1$ must be kept. At order $n=2$ in the derivative expansion, however, this is no longer necessary. Indeed, for $d=4$ the factor $e^{4A}$ coming from the volume element $\sqrt{\g}$ exactly 
cancels the factor $e^{-4A}$ coming from the four-derivative terms. Hence, the divergences at order $n=2$ are only power-like or logarithmic in 
$\l$. The first few terms of the asymptotic form of the boundary terms listed in Table \ref{IHQCD-order-2}, therefore, 
suffice.   
\begin{table}
\[\begin{array}{|c|l|l|}
\hline\hline
I &  \ct^I_1 &\cp_1^I(\l)\\ \hline  & & \\

1 & R &  -2\X(\l)=\frac{\ell}{d-2}\left(1-\frac{\x^2b_0}{d-1}\l+\co(\l^2)\right)\\&&\\

2 & \x^2\l^{-2}\pa_i\l\pa^i\l & -M(\l)=-\frac{\ell}{d-2}\left(1+\left(\frac{2}{d-2}-\frac{\x^2}{d-1}\right)b_0\l+\co(\l^2)\right)\\ &&\\

3 & \pa_i\c\pa^i\c &  -\Th(\l)=-\frac{\ell}{d-2}\left(M_{pl}^3N_c^2\right)^{-1}\left(Z_0+\left(Z_1-\frac{\x^2b_0}{d-1}Z_0\right)\l+
\co(\l^2)\right)\\&&\\

\hline\hline
\end{array}\]
\caption{The leading form of the solution (\ref{solution-parameterization}) of the Hamilton-Jacobi equation at order $n=1$ 
 for IHQCD. The third column gives the asymptotic form of the terms in the fourth column of Table \ref{order-1}.}
\label{IHQCD-order-1}
\end{table}

\begin{landscape}
\begin{table}
\[\begin{array}{|c|l|l|}
\hline\hline
I &  \ct^I_2 &\cp^I_2(\l) \\
\hline & &\\
1 & R^{ij}R_{ij} & \frac{\ell^3}{(d-2)^2b_0}\left(\l^{-1}+\left(\frac{2\x^2b_0}{d-1}-\frac{2U_2}{\x^2b_0}\right)\log\l+\co(\l^0)\right)\\&&\\

2 & R^2 & -\frac{d}{4(d-1)}\frac{\ell^3}{(d-2)^2b_0}\left(\l^{-1}+\left(\frac{2\x^2b_0}{d-1}-\frac{2U_2}{\x^2b_0}\right)\log\l+\co(\l^0)\right)\\ &&\\

3 & \x^2\left(\l^{-2}D^iD^j\l D_iD_j\l-2\l^{-3}\pa_i\l\pa_j\l D^iD^j\l+\l^{-4}(\pa_i\l\pa^i\l)^2\right) & 
    -\frac{\ell^3\x^2b_0}{3(d-1)^2(d-1)^2}\left(\l+\co(\l^2)\right) \\&&\\

4 & \x^3\left(\l^{-3}\pa_i\l\pa_j\l D^iD^j\l-\l^{-4}(\pa_i\l\pa^i\l)^2\right)  & -\frac{2\x^{-1}\ell^3}{3(d-2)^2b_0}\left(\l^{-1}+\co(\l^0)\right) \\&&\\

5 & \x^3\left(\l^{-3}\pa_i\l\pa^i\l \square\l-\l^{-4}(\pa_i\l\pa^i\l)^2\right)  &  \frac{2\x^{-1}\ell^3}{3(d-2)^2b_0}\left(\l^{-1}+\co(\l^0)\right)\\&&\\

6 & \x^2\left(\l^{-2}(\square\l)^2 -2\l^{-3}\pa_i\l\pa^i\l \square\l+\l^{-4}(\pa_i\l\pa^i\l)^2\right) & 
    -\frac{\ell^3}{(d-2)^2b_0}\left(\l^{-1}+\co(\l^0)\right)\\&&\\

7 & \x^4\l^{-4}\left(\pa_i\l\pa^i\l\right)^2  &  -\frac{(3d-4)\ell^3}{12(d-2)^2(d-1)b_0}\left(\l^{-1}+\co(\l^0)\right)\\&&\\

8 & \left(\pa_i\c\pa^i\c\right)^2  & \frac{(3d-4)\ell^3(M_{pl}l^3N_c^2)^{-2}}{4(d-1)(d-2)^2b_0}
       \left(Z_0^2\l^{-1}+\left(\left(\frac{2\x^2b_0}{d-1}-\frac{2U_2}{\x^2b_0}\right)Z_0-2Z_1\right)Z_0\log\l+\co(\l^0)\right)\\&&\\

9 & \x R^{ij}(\l^{-1}D_iD_j\l-\l^{-2}\pa_i\l\pa_j\l) &  -\frac{2\x\ell^3}{(d-1)(d-2)^2}+\co(\l)\\&&\\

10 & \x^2\l^{-2}R^{ij}\pa_i\l\pa_j\l  &  \frac{2\ell^3}{(d-2)^2b_0}\left(\l^{-1}+\co(\l^0)\right)\\&&\\

 \hline\hline
\end{array}\]
\end{table}
\begin{table}
\[\begin{array}{|c|l|l|}
\hline
\hline & & \\

11 & \x R(\l^{-1}\square\l-\l^{-2}\pa_i\l\pa^i\l)  & \frac{2\x\ell^3}{(d-2)^2(d-1)}+\co(\l)\\&&\\

12 & \x^2\l^{-2}R\pa_i\l\pa^i\l  & -\frac{d\ell^3}{2(d-2)^2(d-1)b_0}\left(\l^{-1}+\co(\l^0)\right)\\&&\\

13 & R^{ij}\pa_i\c\pa_j\c  & -\frac{2\ell^3(M_{pl}^3N_c^2)^{-1}}{(d-2)^2b_0}\left(Z_0\l^{-1}+
      \left(\left(\frac{2\x^2b_0}{d-1}-\frac{2U_2}{\x^2b_0}\right)Z_0-Z_1\right)\log\l+\co(\l^0)\right)\\&&\\

14 & \x (\l^{-1}D^iD^j\l\pa_i\c\pa_j\c-\l^{-2}(\pa_i\l\pa^i\c)^2) &  \frac{2\x\ell^3(M_{pl}^3N_c^2)^{-1}}{(d-2)^2(d-1)}Z_0+\co(\l)\\&&\\

15 & \x^2\l^{-2}\left(\pa_i\l\pa^i\c\right)^2 &  -\frac{2\ell^3(M_{pl}^3N_c^2)^{-1}}{(d-2)^2b_0}Z_0\left(\l^{-1}+\co(\l^0)\right)\\&&\\

16 & \x(\l^{-1}\square\l-\l^{-2}\pa_j\l\pa^j\l)\pa_i\c\pa^i\c  & -\frac{\x^{-1}\ell^3(M_{pl}^3N_c^2)^{-1}}{(d-2)^2}
       \left(\frac{2}{d-2}-\frac{2\x^2}{d-1}\right)Z_0+\co(\l)\\&&\\

17 & R\pa_i\c\pa^i\c  &  \frac{d\ell^3(M_{pl}^3N_c^2)^{-1}}{2(d-2)^2(d-1)b_0}\left(Z_0\l^{-1}+
      \left(\left(\frac{2\x^2b_0}{d-1}-\frac{2U_2}{\x^2b_0}\right)Z_0-Z_1\right)\log\l+\co(\l^0)\right)\\&&\\

18 & \x^2\l^{-2}\pa_i\l\pa^i\l\pa_j\c\pa^j\c &  -\frac{d\ell^3(M_{pl}^3N_c^2)^{-1}}{2(d-2)^2(d-1)b_0}Z_0\left(\l^{-1}+\co(\l^0)\right)\\&&\\

19 & \left(\square_\g\c\right)^2  & \frac{\ell^3(M_{pl}^3N_c^2)^{-1}}{(d-2)^2b_0}\left(Z_0\l^{-1}+
      \left(\left(\frac{2\x^2b_0}{d-1}-\frac{2U_2}{\x^2b_0}\right)Z_0-Z_1\right)\log\l+\co(\l^0)\right)\\&&\\

20 & \x\l^{-1}\square_\g\c \pa_i\l\pa^i\c &  -\frac{\x^{-1}\ell^3(M_{pl}^3N_c^2)^{-1}}{(d-2)^2b_0}
      \left(Z_1-\frac{\x^2b_0}{d-1}\right)Z_0+\co(\l)\\
&&\\ \hline\hline
\end{array}\]
\caption{The leading form of the solution (\ref{solution-parameterization}) of the Hamilton-Jacobi equation at order $n=2$ and 
$d=4$ for IHQCD. The third column gives the asymptotic form of the terms in the fourth column of Table \ref{order-2}.}
\label{IHQCD-order-2}
\end{table}
\end{landscape}

The entire boundary term that renders the variational problem for IHQCD well defined, therefore, takes the form
\be\label{IHQCD-counterterms}
\framebox[6.0in]{\rule[-0.9in]{0.0in}{0.0in}
\begin{minipage}{6.5in}
\begin{eqnarray*}
S_{b}^{\rm IHQCD}&=&-\left(\cs\sub{0}+\cs\sub{2}+\widetilde{\cs}\sub{4}\right)\NO\\
&=& -\frac{1}{\k^2}\int d^4x\sqrt{\g}\left(\rule{0in}{0.0in}U(\l)+\frac12M(\l)\x^2\l^{-2}\pa_i\l\pa^i\l+\X(\l)R+\frac12\Th(\l)\pa_i\c\pa^i\c\right.\NO\\
&&\left.+\frac14a(\l)\left(R_{ij}R^{ij}-\frac{1}{3}R^2\right)+\frac14b(\l)
\left[\left(R^{ij}-\frac{1}{3}R\g^{ij}\right)\pa_i\c\pa_j\c-\frac12(\square_\g\c)^2\right]\right.\NO\\
&&\left.+\frac14c(\l)
\left(\pa_i\c\pa^i\c\right)^2\right),
\end{eqnarray*}
\end{minipage}
}
\ee
where we have set $d=4$ and introduced functions $a(\l)$, $b(\l)$ and $c(\l)$, which are given by 
\bea
&&a(\l)=-\frac{\ell^3}{2}\left(b_0^{-1}\l^{-1}+2\z\log\l+c_1\right),\NO\\
&&b(\l)=\ell^3(M_{pl}^3N_c^2)^{-1}Z_0\left(b_0^{-1}\l^{-1}+\left(2\z -\frac{Z_1}{b_0Z_0}\right)\log\l+c_2\right),\NO\\
&&c(\l)=-\frac{\ell^3}{3}(M_{pl}^3N_c^2)^{-2}Z_0^2\left(b_0^{-1}\l^{-1}+\left(2\z-\frac{2Z_1}{b_0Z_0}\right)\log\l+c_3\right).
\eea
Here 
\be
\z=\frac{\x^2}{3}-\frac{U_2}{\x^2b_0^2}.
\ee
and $c_1$, $c_2$ and $c_3$ are arbitrary constants. The terms they multiply correspond to {\em finite} local counterterms and they 
reflect the usual scheme dependence. The functional derivative of each of these three terms w.r.t the induced metric gives 
a local transverse and traceless tensor. These three tensors are given in (\ref{traceless-tensors-1})-(\ref{traceless-tensors-3}) below. 
Moreover, the tilde in the $n=2$ solution, $\cs\sub{4}$, of the Hamilton-Jacobi equation is there to remind us that this is not the full 
solution of the Hamilton-Jacobi equation at order $n=2$. Namely, $\widetilde\cs\sub{4}$ contains only the local divergent part (plus
scheme dependence) of the $n=2$ solution. In addition, there is a {\em finite part}, $\widehat\cs\sub{4}$, that corresponds to 
the renormalized action and which we have not determined. Recall that all the homogeneous solutions of the equations in the 
previous section were ignored precisely because they contribute only to this finite part, which cannot be determined from the 
derivative expansion of the Hamilton-Jacobi equation alone. As usual, to determine this part one must impose some regularity or
boundary condition in the deep interior of the spacetime.   

The form of this boundary term deserves some close inspection. By far the most remarkable feature is that
it is totally covariant. In particular, contrary to the boundary term (\ref{counterterms}) for the strictly
asymptotically locally AdS dilaton-axion system, it {\em does not} break the bulk diffeomorphisms corresponding 
to shifts of the radial coordinate. A direct consequence of this fact is that, as we shall show below, the trace
of the stress tensor does not depend {\em explicitly} on the sources, but only implicitly through the dilaton 
one-point function. In other words, there is no conformal anomaly in IHQCD, although conformal invariance is 
broken via the dilaton one-point function. A related observation is that there are terms proportional to $\l^{-1}\sim r$ at 
order $n=2$ in the derivative expansion. These terms are in fact nothing but the conformal anomaly for a gravity-axion system 
in asymptotically AdS space, as can be seen from (\ref{counterterms}) in Appendix \ref{constant-potential}. This picture
suggests that the field $\l$ effectively acts as a compensator for scale transformations or shifts of the radial 
coordinate\footnote{We are grateful to Kostas Skenderis for pointing out this interpretation to us.}. Later on we will 
confirm this by showing that the source of the dilaton $\l$ can be removed by a bulk diffeomorphism that induces a Weyl 
rescaling of the boundary metric, while the one-point function of the dilaton contains exactly the would-be 
conformal anomaly of the gravity-axion system.

\subsection{One-point functions and Fefferman-Graham expansions}

Having solved the Hamilton-Jacobi equation for IHQCD asymptotically, we are in a position to {\em derive} the 
full asymptotic behavior of the bulk fields, i.e. the generalized Fefferman-Graham expansions, without 
solving asymptotically the second order equations of motion. Not only is this approach considerably more efficient 
than directly solving the second order equations of motion, but it also avoids the necessity of making an ansatz for the
form of the asymptotic expansions. As we shall see, the structure of these asymptotic expansions can be highly involved
and practically impossible to guess a priori. However, the asymptotic solution of the Hamilton-Jacobi equation that we
have obtained above already contains all information about the form of the asymptotic expansions. Additionally, it will
automatically tell us how the one-point functions, i.e. the renormalized momenta, are related to the coefficients of
the asymptotic expansions.  

The key ingredient in deriving the asymptotic expansions from the asymptotic solution of the Hamilton-Jacobi equation is
the first order flow equations
\bea\label{flow-eqs}
&&\dot{\g}_{ij}=4\k^2\left(\g_{ik}\g_{jl}-\frac13\g_{kl}\g_{ij}\right)\frac{1}{\sqrt{\g}}\frac{\d\cs}{\d\g_{kl}},\NO\\
&&\dot{\l}=\k^2\x^{-2}\l^2\frac{1}{\sqrt{\g}}\frac{\d\cs}{\d\l},\NO\\
&&\dot{\c}=\k^2 Z^{-1}(\l)\frac{1}{\sqrt{\g}}\frac{\d\cs}{\d\c},
\eea
which follow from combining the two expressions for the canonical momenta given in (\ref{momenta-scalars}) and (\ref{HJ-momenta}).
Now, the solution of the Hamilton-Jacobi equation we have determined above takes the form   
\be
\cs=\cs\sub{0}+\cs\sub{2}+\widetilde{\cs}\sub{4}+\widehat{\cs}\sub{4},
\ee
where $\widehat{\cs}\sub{4}$ is undetermined and remains finite when the radial cut-off is removed, while the other 
terms are given in (\ref{IHQCD-counterterms}). Inserting this form of the solution in (\ref{flow-eqs}) we obtain 
the following flow equations.  
\bea
\dot{\l}&=& \x^{-2}\l^2\pa_\l U\NO\\
&&+\frac12\l^{-1}\left(2M-\l\pa_\l M\right)\pa_i\l\pa^i\l-M\square_\g\l+\frac12\x^{-2}\l^2\pa_\l\Th \pa_i\c\pa^i\c
+\x^{-2}\l^2 \pa_\l\X R\NO\\
&&+\frac{1}{4\x^2}\l^2\left(\pa_\l a(R_{ij}R^{ij}-\frac13R^2)+\pa_\l b\left[(R_{ij}-\frac13R\g_{ij})\pa^i\c\pa^j\c-\frac12(\square_\g\c)^2\right]+\pa_\l c(\pa_i\c\pa^i\c)^2\right)\NO\\
&&+\k^2\x^{-2}\l^2\frac{1}{\sqrt{\g}}\widehat{\p}_\l\sub{4},
\eea

\bea 
\dot{\c}&=& -Z^{-1}\left(\Th\square_\g\c+\pa_\l\Th \pa_i\l\pa^i\c\right)\NO\\
&&-\frac12Z^{-1}D^i\left(b(R_{ij}-\frac13R\g_{ij})\pa^j\c+\frac12D_i(b\square_\g\c)+2c\pa_k\c\pa^k\c\pa_i\c\right)\NO\\
&&+\k^2Z^{-1}\frac{1}{\sqrt{\g}}\widehat{\p}_\c\sub{4},
\eea

\bea
\dot{\g}_{ij}&=&-\frac23U\g_{ij}\NO\\
&&-4\X R_{ij}+4\pa_\l\X D_iD_j\l+4\left(\pa_\l^2\X-\frac12\x^2\l^{-2}M\right)\pa_i\l\pa_j\l-2\Th\pa_i\c\pa_j\c\NO\\
&&+\frac23\g_{ij}\left(\X R+\frac12M\l^{-2}\x^2\pa_k\l\pa^k\l+\frac12\Th\pa_k\c\pa^k\c\right)\NO\\
&&-2aR_{ikjl}R^{kl}+\frac23aRR_{ij}+2D_{(i}D_k(aR^{k}{}_{j)})-\square_\g(aR_{ij})-\frac23D_iD_j(aR)\NO\\
&&+\g_{ij}\left(\frac12a(R_{kl}R^{kl}-\frac13R^2)-\frac13D_kD_l(aR^{kl})+\frac13\square_\g(aR)\right)\NO\\
&&+\frac12b\g_{ij}(R_{kl}-\frac13R\g_{kl})\pa^k\c\pa^l\c-bR_{(ik}\pa^k\c\pa_{j)}\c+\frac13bR\pa_i\c\pa_j\c-bR_{ikjl}\pa^k\c\pa^l\c\NO\\
&&+\frac13bR_{ij}\pa_k\c\pa^k\c+D_{(i}D^k(b\pa_k\c\pa_{j)}\c)-\frac12\square_\g(b\pa_i\c\pa_j\c)-\frac16\g_{ij}D_kD_l(b\pa^k\c\pa^l\c)\NO\\
&&-\frac13D_iD_j(b\pa_k\c\pa^k\c)+\frac16\g_{ij}\square_\g(b\pa_k\c\pa^k\c)+bD_iD_j\c\square_\g\c-D_{(i}(b\square_\g\c D_{j)}\c)\NO\\
&&+\frac16\g_{ij}D_k(b\square_\g\c D^k\c)-\frac14\g_{ij}b(\square_\g\c)^2
+c\left(\frac12\g_{ij}\pa_k\c\pa^k\c-2\pa_i\c\pa_j\c\right)\pa_l\c\pa^l\c\NO\\
&&+4\k^2\left(\g_{ik}\g_{jl}-\frac13\g_{kl}\g_{ij}\right)\frac{1}{\sqrt{\g}}\widehat{\p}\sub{4}^{kl}.
\eea
Here, $\widehat{\p}_\l\sub{4}$, $\widehat{\p}_\c\sub{4}$ and $\widehat{\p}\sub{4}^{kl}$ denote respectively 
the functional derivatives of $\widehat{\cs}\sub{4}$ w.r.t. the dilaton, the axion and the induced metric. Since
$\widehat{\cs}\sub{4}$ is undetermined, so are these renormalized momenta, which define the one-point functions
of the dual operators \cite{Papadimitriou:2004ap}. The presence of these terms in the flow equation will lead directly 
to the identification of the normalizable modes in the asymptotic expansions. 

Although these flow equations look rather complicated, they can actually be solved in a fairly straightforward way
by noticing that the asymptotic solutions will in fact be two-scale expansions. Namely, there is an expansion in 
exponentials of $r$ coming from the derivative expansion. Moreover, at each order in this expansion there is an 
expansion in powers and possibly logarithms of $r$. Since the structure of the expansion in exponentials of $r$ is 
pretty clear from the derivative expansion, it is useful to separate the two expansions by writing explicitly
\bea
&&\g_{ij}(r,x)=e^{2r/\ell}\left(\g\sub{0}_{ij}(r,x)+e^{-2r/\ell}\g\sub{2}_{ij}(r,x)+e^{-4r/\ell}\g\sub{4}_{ij}(r,x)+\cdots\right),\NO\\
&&\l(r,x)=\l\sub{0}(r,x)+e^{-2r/\ell}\l\sub{2}(r,x)+e^{-4r/\ell}\l\sub{4}(r,x)+\cdots,\NO\\
&&\c(r,x)=\c\sub{0}(r,x)+e^{-2r/\ell}\c\sub{2}(r,x)+e^{-4r/\ell}\c\sub{4}(r,x)+\cdots,
\eea
where the coefficients of the exponentials here are undetermined functions of $r$, only constrained by the requirement 
that their asymptotic expansions only contain powers and logarithms of $r$, but not exponentials. 

Inserting these expansions into the flow equations leads to non-linear first order equations for the order zero 
coefficients and {\em linear} first order equations for the higher order coefficients. Namely,
for the induced metric we get   
\be
\dot{\g}\sub{0}_{ij}+2\left(\frac{1}{\ell}+\frac13 U(\l\sub{0})\right)\g\sub{0}_{ij}=0,
\ee

\bea
\dot{\g}\sub{2}_{ij}+\frac23 U(\l\sub{0})\g\sub{2}_{ij}&=&-4\X(\l\sub{0})R_{ij}[\g\sub{0}]+4\X'(\l\sub{0})D\sub{0}_iD\sub{0}_j\l\sub{0}
-2\Th(\l\sub{0})\pa_i\c\sub{0}\pa_j\c\sub{0}\NO\\
&&+4\left(\X''(\l\sub{0})-\frac12\x^2\l\sub{0}^{-2}M(\l\sub{0})\right)\pa_i\l\sub{0}\pa_j\l\sub{0}\NO\\
&&+\frac23\g\sub{0}_{ij}\left(-U'(\l\sub{0})\l\sub{2}+\X(\l\sub{0})R[\g\sub{0}]\right.\\
&&\left.+\frac12\Th(\l\sub{0})D\sub{0}_k\c\sub{0}D\sub{0}^k\c\sub{0}
+\frac{\x^2}{2}\l\sub{0}^{-2}M(\l\sub{0})D\sub{0}^k\l\sub{0}D\sub{0}_k\l\sub{0}\right),\NO
\eea

\bea
&&\dot{\g}\sub{4}_{ij}+\left(\frac23U-\frac2\ell\right)\g\sub{4}_{ij}=
-\frac23\left(\l\sub{4}U'+\frac12\l\sub{2}^2U''\right)\g\sub{0}_{ij}-\frac23\l\sub{2}U'\g\sub{2}_{ij}\NO\\
&&-4\l\sub{2}\X'R_{ij}[\g\sub{0}]
-4\X\left(D\sub{0}_kD\sub{0}_{(i}\g\sub{2}^k_{j)}-\frac12\square\sub{0}\g\sub{2}_{ij}-\frac12D\sub{0}_iD\sub{0}_j\g\sub{2}\right)\NO\\
&&+4\l\sub{2}\X''D\sub{0}_iD\sub{0}_j\l\sub{0}+4\X'\left(D\sub{0}_iD\sub{0}_j\l\sub{2}
+\left(\frac12D\sub{0}_k\g\sub{2}_{ij}-D\sub{0}_{(i}\g\sub{2}_{kj)}\right)D\sub{0}^k\l\sub{0}\right)\NO\\
&&+2\left(\X''-\frac{\x^2}{2}\l\sub{0}^{-2}M\right)\pa_{(i}\l\sub{0}\pa_{j)}\l\sub{2}
+4\l\sub{2}\left(\X'''-\frac{\x^2}{2}\l\sub{0}^{-2}\left(M'-2\l\sub{0}^{-1}M\right)\right)\pa_i\l\sub{0}\pa_j\l\sub{0}\NO\\
&&-2\l\sub{2}\Th'\pa_i\c\sub{0}\pa_j\c\sub{0}-\Th\pa_{(i}\c\sub{0}\pa_{j)}\c\sub{2}\NO\\
&&+\frac23\g\sub{2}_{ij}\left(\X R[\g\sub{0}]+\frac{\x^2}{2}\l\sub{0}^{-2}MD\sub{0}_k\l\sub{0}D\sub{0}^k\l\sub{0}
+\frac12\Th D\sub{0}_k\c\sub{0}D\sub{0}^k\c\sub{0}\right)\NO\\
&&+\frac23\g\sub{0}_{ij}\left[\l\sub{2}\X' R[\g\sub{0}]+\X\left(D\sub{0}^kD\sub{0}^l\g\sub{2}_{kl}-\square\sub{0}\g\sub{2}
-\g\sub{2}^{kl}R[\g\sub{0}]_{kl}\right)+\frac{\x^2}{2}\l\sub{0}^{-2}\times\right.\NO\\
&&\left.\left(\l\sub{2}(M'-2\l\sub{0}^{-1}M)D\sub{0}_k\l\sub{0}D\sub{0}^k\l\sub{0}+2MD\sub{0}_k\l\sub{0}D\sub{0}^k\l\sub{2}
-M\g\sub{2}^{kl}\pa_k\l\sub{0}\pa_l\l\sub{0}\right)\right.\NO\\
&&\left.+\frac12\l\sub{2}\Th' D\sub{0}_k\c\sub{0}D\sub{0}^k\c\sub{0}+\Th  D\sub{0}_k\c\sub{0}D\sub{0}^k\c\sub{2}
-\frac12\Th\g\sub{2}^{kl}\pa_k\c\sub{0}\pa_l\c\sub{0}\right]\NO\\
&&-2aR_{ikjl}[\g\sub{0}]R^{kl}[\g\sub{0}]+\frac23aR[\g\sub{0}]R_{ij}[\g\sub{0}]+2D\sub{0}_{(i}D\sub{0}_k(aR^{k}{}_{j)}[\g\sub{0}])
-\square\sub{0}(aR_{ij}[\g\sub{0}])\NO\\
&&-\frac23D\sub{0}_iD\sub{0}_j(aR[\g\sub{0}])+\g\sub{0}_{ij}\left(\frac12a(R_{kl}[\g\sub{0}]R^{kl}[\g\sub{0}]-\frac13R^2[\g\sub{0}])
-\frac13D\sub{0}_kD\sub{0}_l(aR^{kl}[\g\sub{0}])\right.\NO\\
&&\left.+\frac13\square\sub{0}(aR[\g\sub{0}])\right)
+\frac12b\g\sub{0}_{ij}\left(R_{kl}[\g\sub{0}]-\frac13R[\g\sub{0}]\g\sub{0}_{kl}\right)D\sub{0}^k\c\sub{0}D\sub{0}^l\c\sub{0}\NO\\
&&-bR_{(ik}[\g\sub{0}]D\sub{0}^k\c\sub{0}D\sub{0}_{j)}\c\sub{0}+\frac13bR[\g\sub{0}]\pa_i\c\sub{0}\pa_j\c\sub{0}
-bR_{ikjl}[\g\sub{0}]D\sub{0}^k\c\sub{0}D\sub{0}^l\c\sub{0}\NO\\
&&+\frac13bR_{ij}[\g\sub{0}]D\sub{0}_k\c\sub{0}D\sub{0}^k\c\sub{0}
-\frac16\g\sub{0}_{ij}D\sub{0}_kD\sub{0}_l(bD\sub{0}^k\c\sub{0}D\sub{0}^l\c\sub{0})\NO\\
&&-\frac12\square\sub{0}(b\pa_i\c\sub{0}\pa_j\c\sub{0})+D\sub{0}_{(i}D\sub{0}^k(b\pa_k\c\sub{0}\pa_{j)}\c\sub{0})-\frac13D\sub{0}_iD\sub{0}_j(bD\sub{0}_k\c\sub{0}D\sub{0}^k\c\sub{0})\NO\\
&&+\frac16\g\sub{0}_{ij}\square\sub{0}(bD\sub{0}_k\c\sub{0}D\sub{0}^k\c\sub{0})+bD\sub{0}_iD\sub{0}_j\c\sub{0}\square\sub{0}\c\sub{0}
-D\sub{0}_{(i}(b\square\sub{0}\c\sub{0} D\sub{0}_{j)}\c\sub{0})\NO\\
&&+\frac16\g\sub{0}_{ij}D\sub{0}_k(b\square\sub{0}\c\sub{0} D\sub{0}^k\c\sub{0})-\frac14\g\sub{0}_{ij}b(\square\sub{0}\c\sub{0})^2\NO\\
&&+c\left(\frac12\g\sub{0}_{ij}D\sub{0}_k\c\sub{0}D\sub{0}^k\c\sub{0}-2\pa_i\c\sub{0}\pa_j\c\sub{0}\right)D\sub{0}_l\c\sub{0}D\sub{0}^l\c\sub{0}\NO\\
&&+4\k^2\left(\g\sub{0}_{ik}\g\sub{0}_{jl}-\frac13\g\sub{0}_{kl}\g\sub{0}_{ij}\right)\frac{1}{\sqrt{\g\sub{0}}}\widehat{\p}\sub{4}^{kl}.
\eea

Similarly, the first order equations for the coefficients in the dilaton expansion are 
\be
\dot{\l}\sub{0}-\x^{-2}\l^2_{(0)}U'(\l\sub{0})=0,
\ee

\bea
&&\dot{\l}\sub{2}-\left(\frac{2}{\ell}+2\x^{-2}\l\sub{0}U'(\l\sub{0})+\x^{-2}\l\sub{0}^2U''(\l\sub{0})\right)\l\sub{2}=
\x^{-2}\l\sub{0}^2\X'(\l\sub{0})R[\g\sub{0}]\NO\\
&&-M(\l\sub{0})\square\sub{0}\l\sub{0}+\frac12\l\sub{0}^{-1}\left(2M(\l\sub{0})
-\l\sub{0}M'(\l\sub{0})\right)D\sub{0}_i\l\sub{0}D\sub{0}^i\l\sub{0}\NO\\
&&+\frac12\x^{-2}\l\sub{0}^2\Th'(\l\sub{0})D\sub{0}_i\c\sub{0}D\sub{0}^i\c\sub{0},
\eea

\bea
&&\dot{\l}\sub{4}-\left(\frac{4}{\ell}+2\x^{-2}\l\sub{0}U'+\x^{-2}\l\sub{0}^2U''\right)\l\sub{4}=
\left(U'+2\l\sub{0}U''+\frac12\l\sub{0}^2U'''\right)\x^{-2}\l\sub{2}^2\NO\\
&&-\l\sub{2}M'\square\sub{0}\l\sub{0}+M\left(\g\sub{2}^{ij}D\sub{0}_iD\sub{0}_j\l\sub{0}
-\square\sub{0}\l\sub{2}+\frac12(2D\sub{0}_i\g\sub{2}^i_j-D\sub{0}_j\g\sub{2})D\sub{0}^j\l\sub{0}\right)\NO\\
&&+\frac{\x^{-2}}{2}\l\sub{0}\l\sub{2}\left(2\Th'+\l\sub{0}\Th''\right)D\sub{0}_i\c\sub{0}D\sub{0}^i\c\sub{0}
+\x^{-2}\l\sub{0}\l\sub{2}\left(2\X'+\l\sub{0}\X''\right)R[\g\sub{0}]\NO\\
&&+\frac{\x^{-2}}{2}\l\sub{0}^2\Th'\left(2D\sub{0}^i\c\sub{0}D\sub{0}_i\c\sub{2}-\g\sub{2}^{ij}\pa_i\c\sub{0}\pa_j\c\sub{0}\right)\NO\\
&&+\x^{-2}\l\sub{0}^2\X'\left(D\sub{0}^iD\sub{0}^j\g\sub{2}_{ij}-\square\sub{0}\g\sub{2}-\g\sub{2}^{ij}R[\g\sub{0}]_{ij}\right)\NO\\
&&+\frac12\l\sub{2}\l\sub{0}^{-1}\left(M'-\l\sub{0}M''\right)D\sub{0}_i\l\sub{0}D\sub{0}^i\l\sub{0}
+\frac12\l\sub{0}^{-1}\left(M'-2\l\sub{0}^{-1}M\right)\times\NO\\
&&\left(\l\sub{2}D\sub{0}_i\l\sub{0}D\sub{0}^i\l\sub{0}
+\l\sub{0}\g\sub{2}^{ij}\pa_i\l\sub{0}\pa_j\l\sub{0}-2\l\sub{0}D\sub{0}_i\l\sub{0}D\sub{0}^i\l\sub{2}\right)\NO\\
&&+\frac{1}{4\x^2}\l\sub{0}^2\left(a'\left(R_{ij}[\g\sub{0}]R^{ij}[\g\sub{0}]-\frac13R^2[\g\sub{0}]\right)
+c'\left(D\sub{0}_i\c\sub{0}D\sub{0}^i\c\sub{0}\right)^2\right.\NO\\
&&\left.+b'\left[\left(R_{ij}[\g\sub{0}]-\frac13R[\g\sub{0}]\g\sub{0}_{ij}\right)D\sub{0}^i\c\sub{0}D\sub{0}^j\c\sub{0}
-\frac12(\square\sub{0}\c\sub{0})^2\right]\right)\NO\\
&&+\k^2\x^{-2}\l\sub{0}^2\frac{1}{\sqrt{\g\sub{0}}}\widehat{\p}_\l\sub{4}.
\eea

Finally, for the axion we get
\be
\dot{\c}\sub{0}=0,
\ee

\bea
\dot{\c}\sub{2}-\frac{2}{\ell}\c\sub{2}=-Z^{-1}(\l\sub{0})\left(\Th(\l\sub{0})\square\sub{0}\c\sub{0}+\Th'(\l\sub{0})D\sub{0}_i\l\sub{0}D\sub{0}^i\c\sub{0}\right),
\eea

\bea
&&\dot{\c}\sub{4}-\frac4\ell\c\sub{4}=-\l\sub{2}Z^{-1}\left(\Th'-Z^{-1}Z'\Th\right)\square\sub{0}\c\sub{0}
+Z^{-1}\Th\g\sub{2}^{ij}D\sub{0}_iD\sub{0}_j\c\sub{0}\NO\\
&&-Z^{-1}\Th\left(\square\sub{0}\c\sub{2}-\frac12\left(2D\sub{0}_i\g\sub{2}^i_j-D\sub{0}_j\g\sub{2}\right)D\sub{0}^j\c\sub{0}\right)\NO\\
&&-Z^{-1}\Th'\left(D\sub{0}_{i}\c\sub{0}D\sub{0}^i\l\sub{2}+D\sub{0}_{i}\c\sub{2}D\sub{0}^i\l\sub{0}-\g\sub{2}^{ij}\pa_i\l\sub{0}\pa_j\c\sub{0}\right)\NO\\
&&-Z^{-1}\l\sub{2}\left(\Th''-Z^{-1}Z'\Th'\right)D\sub{0}_i\l\sub{0}D\sub{0}^i\c\sub{0}\NO\\
&&-\frac12Z^{-1}D\sub{0}^i\left(b\left(R_{ij}[\g\sub{0}]-\frac13R[\g\sub{0}]\g\sub{0}_{ij}\right)D\sub{0}^j\c\sub{0}
+\frac12D\sub{0}_i(b\square\sub{0}\c\sub{0})\right.\NO\\
&&\left.+2cD\sub{0}_k\c\sub{0}D\sub{0}^k\c\sub{0}\pa_i\c\sub{0}\rule{0cm}{0.5cm}\right)+\k^2Z^{-1}\frac{1}{\sqrt{\g\sub{0}}}\widehat{\p}_\c\sub{4}.
\eea
In all these first order equations all functions are functions of $\l\sub{0}(r,x)$ and primes denote derivative w.r.t. $\l\sub{0}$. 

From these equations we can now easily construct the asymptotic expansions. The non-linear equations
for the zero order coefficients can be integrated exactly in terms of the function $U(\l\sub{0})$. Using 
the asymptotic form of $U(\l)$ in (\ref{U-UV}) one obtains the asymptotic expansions    
\bea\label{leading-asymptotics}
\g\sub{0}_{ij}(r,x)&=&\left(\frac{r}{\ell}\right)^{2\x^2/3}\left(\rule{.0in}{0.25in}1-\frac{4U_2}{3b_0^2}\left(\frac{\ell}{r}\right)\log(r/\ell)
-\frac23\left(\frac{U_2}{b_0^2}+b_0\x^2\bar{\l}\sub{0}(x)\right)\left(\frac{\ell}{r}\right)\right.\NO\\
&&\left.\rule{.0in}{0.0in}+\frac{4U_\a}{3(\a-2)b_0^\a}\left(\frac{\ell}{r}\right)^{\a-1}+\co\left(\log^2(r/\ell)/r^2\right)\right)\bar{g}\sub{0}_{ij}(x),\\
\l\sub{0}(r,x)&=&\frac{\ell}{b_0 r}+\frac{2U_2}{\x^2b_0^3}\left(\frac{\ell}{r}\right)^2\log(r/\ell)+\left(\frac{\ell}{r}\right)^2\bar{\l}\sub{0}(x)
-\frac{\a U_\a}{(\a-2)\x^2b_0^{\a+1}}\left(\frac\ell r\right)^\a\NO\\
&&+\co\left(\log^2(r/\ell)/r^3\right),\NO\\
\c\sub{0}(r,x)&=&\bar{\c}\sub{0}(x),
\eea
where $\bar{g}\sub{0}_{ij}(x)$, $\bar{\l}\sub{0}(x)$ and $\bar{\c}\sub{0}(x)$ are arbitrary and we identify them
with the {\em sources} of the dual operators. A striking feature of the zero order dilaton expansion is that the source, 
$\bar{\l}\sub{0}(x)$, that couples to the dual operator appears not in the leading order, but at $\co(1/r^2)$. This 
is related to the fact that $\bar{\l}\sub{0}(x)$ can be removed by a bulk diffeomorphism corresponding to 
shifts of the radial coordinate, which we will demonstrate below. However, we have already made extensive use 
of the identity $\pa_i\l=\co(\l^2)$ in deriving the results in Tables \ref{IHQCD-order-1} and \ref{IHQCD-order-2}, 
which follows precisely from the observation that the source of the dilaton appears at subleading order.  

At the second order we get 
\bea
&&\g\sub{2}_{ij}(r,x)=\bar{g}\sub{2}_{ij}(x)+\co\left(\frac1r\log(r/\ell)\right),\NO\\
&&\l\sub{2}(r,x)=\left(\frac{\ell}{r}\right)^{2+2\x^2/3}\left(\bar{\l}\sub{2}(x)+\co\left(\frac{1}{r}\log(r/\ell)\right)\right),\NO\\
&&\c\sub{2}(r,x)= \left(\frac{\ell}{r}\right)^{2\x^2/3}\left(\bar{\c}\sub{2}(x)+\co\left(\frac{1}{r}\log(r/\ell)\right)\right),
\eea
where
\bea
\bar{g}\sub{2}_{ij}(x)&=&-\frac{\ell^2}{2}\left(R_{ij}[\bar{g}\sub{0}]-\frac16\bar{g}\sub{0}_{ij}R[\bar{g}\sub{0}]
\right.\\
&&\left.-\left(M_{pl}^3N_c^2\right)^{-1}Z_0\left(\pa_i\bar{\c}\sub{0}\pa_j\bar{\c}\sub{0}
-\frac16\bar{g}\sub{0}_{ij}\bar{D}\sub{0}_k\bar{\c}\sub{0} \bar{D}\sub{0}^k\bar{\c}\sub{0}\right)
\right),\NO\\
\bar{\l}\sub{2}(x)&=&-\frac{\ell^2}{4}
\left(\frac{1}{6b_0}R[\bar{g}\sub{0}]-\lbar{\square}\sub{0}\bar{\l}\sub{0}\right.\\
&&\left.+\frac{1}{2\x^2 b_0^2}\left(M_{pl}^3N_c^2\right)^{-1}\left(Z_1-\frac{\x^2b_0}{3}Z_0\right)
\bar{D}\sub{0}_k\bar{\c}\sub{0} \bar{D}\sub{0}^k\bar{\c}\sub{0}
\right),\NO\\
\bar{\c}\sub{2}(x)&=& \frac{\ell^2}{4}\left(M_{pl}^3N_c^2\right)^{-1}\lbar{\square}\sub{0}\bar{\c}\sub{0}.
\eea

Finally, at fourth order the asymptotic expansions take the form
\bea
&&\g\sub{4}_{ij}(r,x)=\left(\frac{\ell}{r}\right)^{2\x^2/3}\left(\left(\frac{r}{\ell}\right)\bar{g}\sub{4}_{ij}(x)
+\log(r/\ell)\tilde{g}\sub{4}_{ij}(x)+\hat{g}\sub{4}_{ij}(x)+\co\left(\frac{\log(r/\ell)}{r}\right)\right),\NO\\
&&\l\sub{4}(r,x)=\left(\frac{\ell}{r}\right)^{4\x^2/3}\left(\hat{\l}\sub{4}(x)+\co\left(\frac{\log(r/\ell)}{r}\right)\right),\\
&&\c\sub{4}(r,x)=\left(\frac{\ell}{r}\right)^{4\x^2/3}\left(\left(\frac{r}{\ell}\right)\bar{\c}\sub{4}(x)
+\log(r/\ell)\tilde{\c}\sub{4}(x)+\hat{\c}\sub{4}(x)+\co\left(\frac{\log(r/\ell)}{r}\right)\right),\NO
\eea
where the terms $\hat{g}\sub{4}_{ij}(x)$, $\hat{\l}\sub{4}(x)$ and $\hat{\c}\sub{4}(x)$ are undetermined and are
therefore identified with the normalizable modes. The flow equations relate these to the undetermined renormalized momenta,
thus leading to the expressions for the one-point functions in terms of the coefficients of the asymptotic expansions, which 
we present below. 

In order to write down the explicit expressions for the coefficients in these fourth order expansions, it is useful to define the 
traceless tensors
\bea\label{traceless-tensors-1}
H_{1ij}&=&-2\left(R_{ikjl}-\frac14R_{kl}\bar g\sub{0}_{ij}\right)R^{kl}+\frac23R\left(R_{ij}-\frac14R\bar g\sub{0}_{ij}\right)\NO\\
&&+\frac13\bar{D}\sub{0}_i\bar{D}\sub{0}_jR-\lbar{\square}\sub{0}R_{ij}+\frac16\bar g\sub{0}_{ij}\lbar{\square}\sub{0}R,\\
H_{2ij}&=&\left(-R_{ikjl}+\frac12\bar g\sub{0}_{ij}R_{kl}-R_{(ik}\bar g\sub{0}_{j)l}\right.\NO\\
&&\left.-\frac16R\bar g\sub{0}_{ij}\bar g\sub{0}_{kl}+\frac13R\bar g\sub{0}_{ik}\bar g\sub{0}_{jl}+\frac13R_{ij}\bar g\sub{0}_{kl}\right)
\bar{D}\sub{0}^k\bar{\c}\sub{0}\bar{D}\sub{0}^l\bar{\c}\sub{0}\NO\\
&&+\bar{D}\sub{0}_{(i}\bar{D}\sub{0}^k(\pa_k\bar{\c}\sub{0}\pa_{j)}\bar{\c}\sub{0})-\frac12\lbar{\square}\sub{0}(\pa_i\bar{\c}\sub{0}\pa_j\bar{\c}\sub{0})\NO\\
&&-\frac16\bar g\sub{0}_{ij}\bar{D}\sub{0}_k\bar{D}\sub{0}_l(\bar{D}\sub{0}^k\bar{\c}\sub{0}\bar{D}\sub{0}^l\bar{\c}\sub{0})\NO\\
&&-\frac13\bar{D}\sub{0}_i\bar{D}\sub{0}_j(\bar{D}\sub{0}_k\bar{\c}\sub{0}\bar{D}\sub{0}^k\bar{\c}\sub{0})
+\frac16\bar g\sub{0}_{ij}\lbar{\square}\sub{0}(\bar{D}\sub{0}_k\bar{\c}\sub{0}\bar D\sub{0}^k\bar\c\sub{0})\NO\\
&&+\bar D\sub{0}_i\bar D\sub{0}_j\bar\c\sub{0}\lbar\square\sub{0}\bar\c\sub{0}
-\bar D\sub{0}_{(i}(\lbar\square\sub{0}\bar\c\sub{0} \bar D\sub{0}_{j)}\bar\c\sub{0})\NO\\
&&+\frac16\bar g\sub{0}_{ij}\bar D\sub{0}_k(\lbar\square\sub{0}\bar{\c}\sub{0} \bar D\sub{0}^k\bar\c\sub{0})
-\frac14\bar g\sub{0}_{ij}(\lbar\square\sub{0}\c\sub{0})^2,\label{traceless-tensors-2}\\
H_{3ij}&=&\left(\frac12\bar g\sub{0}_{ij}\bar D\sub{0}_k\bar\c\sub{0}\bar D\sub{0}^k\bar\c\sub{0}-2\pa_i\bar\c\sub{0}\pa_j\bar\c\sub{0}\right)
\bar D\sub{0}_l\bar\c\sub{0}\bar D\sub{0}^l\bar\c\sub{0}.\label{traceless-tensors-3}
\eea
As mentioned in the previous section, these correspond respectively to the derivative of the three terms proportional to 
$c_1$, $c_2$ and $c_3$ in the boundary term (\ref{IHQCD-counterterms}) w.r.t. the induced metric. All curvatures here 
are curvatures of the boundary metric $\bar g\sub{0}_{ij}$. The first two coefficients in the fourth order expansion of 
the metric are just linear combinations of these three traceless tensors. Namely, 
\bea
\bar{g}\sub{4}_{ij}(x)= \frac{\ell^4}{8}H_{1ij}-\frac{\ell^4}{4}Z_0\left(M_{pl}^3N_c^2\right)^{-1}H_{2ij}
+\frac{\ell^4}{12}Z_0^2\left(M_{pl}^3N_c^2\right)^{-2}H_{3ij},
\eea
\bea
\tilde{g}\sub{4}_{ij}(x)&=& \frac{\ell^4}{12}\left(\frac{2U_2}{b_0^2}-\x^2\right)H_{1ij}
-\frac{\ell^4}{4}Z_0\left(M_{pl}^3N_c^2\right)^{-1}\left(\frac{4U_2}{3b_0^2}-\frac{2\x^2}{3}+\frac{Z_1}{b_0Z_0}\right)H_{2ij}\NO\\
&&+\frac{\ell^4}{12}Z_0^2\left(M_{pl}^3N_c^2\right)^{-2}\left(\frac{4U_2}{3b_0^2}-\frac{2\x^2}{3}+\frac{2Z_1}{b_0Z_0}\right)H_{3ij}.
\eea
Moreover, for the coefficients of the axion expansion we have 
\bea
\bar{\c}\sub{4}(x)&=&\frac{\ell^4}{8}\bar D\sub{0}^i\left(\left(M_{pl}^3N_c^2\right)^{-1}
\left[\left(R_{ij}[\bar g\sub{0}]-\frac13R[\bar g\sub{0}]\bar g\sub{0}_{ij}\right)\bar D\sub{0}^j\bar\c\sub{0}
+\frac12D\sub{0}_i\square\sub{0}\c\sub{0}\right]\right.\NO\\
&&\left.-\frac23Z_0\left(M_{pl}^3N_c^2\right)^{-2}\bar D\sub{0}_k\bar\c\sub{0}\bar D\sub{0}^k\bar\c\sub{0}\pa_i\bar\c\sub{0}\right), 
\eea
and
\bea
\tilde{\c}\sub{4}(x)&=&\frac{\ell^4}{8}\bar D\sub{0}^i\left(\left(M_{pl}^3N_c^2\right)^{-1}\left(\frac{8U_2}{3b_0^2}-\frac{2\x^2}{3}+\frac{Z_1}{b_0Z_0}\right)\times\right.\NO\\
&&\left.\left[\left(R_{ij}[\bar g\sub{0}]-\frac13R[\bar g\sub{0}]\bar g\sub{0}_{ij}\right)\bar D\sub{0}^j\bar\c\sub{0}
+\frac12D\sub{0}_i\square\sub{0}\c\sub{0}\right]\right.\NO\\
&&\left.-\frac23Z_0\left(M_{pl}^3N_c^2\right)^{-2}\left(\frac{8U_2}{3b_0^2}-\frac{2\x^2}{3}+\frac{2Z_1}{b_0Z_0}\right)\bar D\sub{0}_k\bar\c\sub{0}\bar D\sub{0}^k\bar\c\sub{0}\pa_i\bar\c\sub{0}\right). 
\eea

Finally, we can write down the general expressions for the one-point functions, i.e. the renormalized momenta, in terms 
of the coefficients of the asymptotic expansions. Starting from the dilaton, the exact one-point function in given by
\bea\label{dilaton-1pt}
\left\langle\co_\l\right\rangle_{ren}&=& -\frac{b_0\ell^3}{8\k^2}\left(\frac{32b_0\x^2}{\ell^{4}}\hat{\l}\sub{4}+
R_{ij}[\bar{g}\sub{0}]R^{ij}[\bar{g}\sub{0}]-\frac13R^2[\bar{g}\sub{0}]\right.\NO\\
&&\left.
-2Z_0\left(M_{pl}^3N_c^2\right)^{-1}\left[\left(R_{ij}[\bar{g}\sub{0}]-\frac13R[\bar{g}\sub{0}]\bar{g}\sub{0}_{ij}\right)
\bar{D}\sub{0}^i\bar{\c}\sub{0}\bar{D}\sub{0}^j\bar{\c}\sub{0}-\frac12(\lbar{\square}\sub{0}\bar{\c}\sub{0})^2\right]\right.\NO\\
&&\left.+\frac23Z_0^2\left(M_{pl}^3N_c^2\right)^{-2}\left(\bar{D}\sub{0}^i\bar{\c}\sub{0}\bar{D}\sub{0}_i\bar{\c}\sub{0}\right)^2\right).
\eea
Notice that this gets contributions from the dilaton normalizable mode, $\hat{\l}\sub{4}$, plus a combination
of the metric and axion sources which is nothing but the conformal anomaly of the axion-gravity system in strictly 
asymptotically AdS space (cf. (\ref{counterterms})). 

All results quoted so far are valid for arbitrary dilaton source $\bar{\l}\sub{0}(x)$. However, in order to 
simplify the expressions for the stress tensor and axion one-point functions, we will only give the expressions for 
{\em constant} $\bar{\l}\sub{0}$, independent of the transverse coordinates. This is not a big disadvantage
since, as we shall see, the dilaton source can be removed or restored by a bulk diffeomorphism corresponding to a Weyl 
transformation of the boundary metric. The full dependence on a generic dilaton source can therefore be restored starting 
from the expressions given below for constant $\bar{\l}\sub{0}(x)$, by a suitable boundary Weyl transformation.       

The one-point function of the stress tensor can be written in the form 
\bea\label{stress-tensor-1pt}
\left\langle T_{ij}\right\rangle_{ren}&=&\frac{2}{\k^2\ell}\left(\Om_{ij}-\Tr\Om \bar g\sub{0}_{ij}\right)-\frac{1}{4b_0}\left\langle\co_\l\right\rangle_{ren}\bar{g}\sub{0}_{ij},
\eea
where the tensor $\Om_{ij}$ is given by
\bea
\Om_{ij}&=&\hat{g}\sub{4}_{ij}+\frac{\x^2b_0}{3}\hat{\l}\sub{4}\bar g\sub{0}_{ij}\NO\\
&&-\frac{\ell^4}{8}c'_1H_{1ij}
+\frac{\ell^4}{4}Z_0\left(M_{pl}^3N_c^2\right)^{-1}
c'_2H_{2ij}-\frac{\ell^4}{12}Z_0^2\left(M_{pl}^3N_c^2\right)^{-2}
c'_3H_{3ij}\NO\\
&&+\frac{\ell^2}{4}\left(\bar D\sub{0}_k\bar D\sub{0}_{(i}\bar g\sub{2}^k_{j)}-\frac12\lbar\square\sub{0}\bar g\sub{2}_{ij}
-\frac12\bar D\sub{0}_i\bar D\sub{0}_j\Tr\bar g\sub{2}\right)\NO\\
&&-\frac{\ell^2}{8}Z_0\left(M_{pl}^3N_c^2\right)^{-1}\pa_{(i}\bar\c\sub{0}\pa_{j)}\bar\c\sub{2}\NO\\
&&-\frac{\ell^2}{24}\bar g\sub{2}_{ij}\left(R+Z_0\left(M_{pl}^3N_c^2\right)^{-1}\bar D\sub{0}_k\bar\c\sub{0}\bar D\sub{0}^k\bar\c\sub{0}\right)\NO\\
&&-\frac{\ell^2}{24}\bar g\sub{0}_{ij}\left(\bar D\sub{0}^k\bar D\sub{0}^l\bar g\sub{2}_{kl}-\lbar\square\sub{0}\Tr\bar g\sub{2}
-\bar g\sub{2}^{kl}R_{kl}\right)\NO\\
&&-\frac{\ell^2}{12}Z_0\left(M_{pl}^3N_c^2\right)^{-1}\bar g\sub{0}_{ij}\left(\bar D\sub{0}^k\bar\c\sub{0}\bar D\sub{0}_k\bar\c\sub{2}
-\frac12\bar g\sub{2}^{kl}\pa_k\bar\c\sub{0}\pa_l\bar\c\sub{0}\right)\NO\\
&&+\frac{\ell^4}{192}\bar g\sub{0}_{ij}\left(R_{kl}R^{kl}-\frac13R^2\right.\NO\\
&&\left.
-2Z_0\left(M_{pl}^3N_c^2\right)^{-1}\left[\left(R_{kl}-\frac13R\bar{g}\sub{0}_{kl}\right)
\bar{D}\sub{0}^k\bar{\c}\sub{0}\bar{D}\sub{0}^l\bar{\c}\sub{0}-\frac12(\lbar{\square}\sub{0}\bar{\c}\sub{0})^2\right]\right.\NO\\
&&\left.+\frac23Z_0^2\left(M_{pl}^3N_c^2\right)^{-2}\left(\bar{D}\sub{0}^i\bar{\c}\sub{0}\bar{D}\sub{0}_i\bar{\c}\sub{0}\right)^2\right).
\eea
Here we have defined the shifted constants 
\bea
&&c'_1=c_1+\frac{2U_2}{3b_0^2}+\left(\frac{2\x^2}{3}-1\right)b_0\bar{\l}\sub{0}-2\z\log b_0+\frac14\left(\frac{4\x^2}{3}-1\right),\NO\\
&&c'_2=c_2+\frac{2U_2}{3b_0^2}+\left(\frac{2\x^2}{3}-1\right)b_0\bar{\l}\sub{0}-\left(2\z-\frac{Z_1}{b_0Z_0}\right)\log b_0+\frac14\left(\frac{4\x^2}{3}-1\right),\NO\\
&&c'_3=c_3+\frac{2U_2}{3b_0^2}+\left(\frac{2\x^2}{3}-1\right)b_0\bar{\l}\sub{0}-\left(2\z-\frac{2Z_1}{b_0Z_0}\right)\log b_0+\frac14\left(\frac{4\x^2}{3}-1\right).\NO\\
\eea
In principle, one can set these constants to zero by suitable choice of scheme, i.e. by a suitable choice of $c_1$, $c_2$ and 
$c_3$, but we give the full expressions so that one knows exactly what scheme needs to be chosen to achieve this. 

Finally, the one-point function of the operator dual to the axion is 
\bea\label{axion-1pt}
\left\langle\co_\c\right\rangle_{ren}&=& -\frac{4Z_0}{\k^2\ell}\hat\c\sub{4}\NO\\
&&+\frac{\ell^3}{2\k^2}Z_0\left(M_{pl}^3N_c^2\right)^{-1}
c''_2\bar D\sub{0}^i\left[\left(R_{ij}-\frac13 R\bar g\sub{0}_{ij}\right)\bar D\sub{0}^j\bar\c\sub{0}+\frac12\bar D\sub{0}_i\lbar\square\sub{0}\bar\c\sub{0}\right]\NO\\
&&-\frac{\ell^3}{3\k^2}Z_0^2\left(M_{pl}^3N_c^2\right)^{-2}
c''_3\bar D\sub{0}^i\left[\bar D\sub{0}_k\bar\c\sub{0}\bar D\sub{0}^k\bar\c\sub{0}\pa_i\bar\c\sub{0}\right]\NO\\
&&+\frac{\ell}{2\k^2}Z_0\left(M_{pl}^3N_c^2\right)^{-1}\times\\
&&\left[\lbar\square\sub{0}\bar\c\sub{2}-\frac12\left(2\bar D\sub{0}_i\bar g\sub{2}^i_j-\bar D\sub{0}_j\Tr\bar g\sub{2}\right)\bar D\sub{0}^j\bar\c\sub{0}
-\bar g\sub{2}^{ij}\bar D\sub{0}_i\bar D\sub{0}_j\bar\c\sub{0}\right],\NO
\eea
where again we have introduced the constants 
\bea
&&c''_2=c'_2+\frac{2U_2}{3b_0^2}+\frac{2\x^2}{3}b_0\bar{\l}\sub{0}-\frac{Z_1}{b_0Z_0},\NO\\
&&c''_3=c'_3+\frac{2U_2}{3b_0^2}+\frac{2\x^2}{3}b_0\bar{\l}\sub{0}-\frac{Z_1}{b_0Z_0},
\eea
to abbreviate the above expression.

\subsection{Asymptotic diffeomorphisms and Ward identities} 

Now that we have determined the general form of the asymptotic expansions for IHQCD and we have 
identified the exact one-point functions, we can proceed with the derivation of the holographic Ward identities. 
These follow as a consequence of the existence of a class of asymptotic bulk diffeomorphisms that 
preserve the structure of the asymptotic expansions.  

Let us consider a generic infinitesimal bulk diffeomorphism, $\d x^\m=-\x^\m$, and demand that it preserves the gauge fixed 
form of the metric, namely that it does not modify the lapse and shift functions. This requirement leads to a pair of equations 
for the vector field $\x^\m$, namely  
\bea
&&\mathcal{L}_\xi g_{rr}=\dot{\xi}^r=0,\NO\\
&&\mathcal{L}_\xi g_{ri}=\g_{ij}(\dot{\xi}^j+\pa^j\xi^r)=0,
\eea
where $\cl_\x$ is the Lie derivative w.r.t. the bulk vector $\x^\m$. Solving these conditions gives
\bea
&&\xi^r=\d\s(x),\NO\\
&&\xi^i=\xi_o^i(x)+\pa_j\d\s(x)\int_r^\infty dr'\g^{ji}(r',x),
\eea
where $\s(x)$ is an arbitrary function of the transverse coordinates and $\xi_o^i(x)$ is an arbitrary transverse vector field. 
Inserting now the asymptotic form of the induced metric we obtain
\be
\xi^i=\xi_o^i(x)+\frac\ell2 e^{-2r/\ell}\left(\frac{\ell}{r}\right)^{\frac{2\x^2}{3}}\bar g\sub{0}^{ij}\pa_j\d\s(x)+
\co\left(e^{-2r/\ell}r^{-\frac{2\x^2}{3}-1}\log(r/\ell)\right).
\ee
Under this bulk diffeomorphism then the induced fields transform as
\bea\label{asymptotic-diffeo-transformations}
&&\d_\x \g_{ij}=\mathcal{L}_\xi g_{ij}=L_\xi\g_{ij}+2K_{ij}\xi^r =L_{\xi_o}\g_{ij}-\frac{2}{d-1}U(\l) \g_{ij}\d\s(x)+\co(r^{-2\x^2/3}),\NO\\
&&\d_\x \l=\mathcal{L}_\xi\l=L_\xi\l+\xi^r\dot{\l}= \xi_o^i\pa_i\l+\x^{-2}\l^2\frac{\pa U}{\pa\l}\d\s(x)+\co (r^{-2-2\x^2/3} e^{-2r/\ell}), \NO\\
&&\d_\x \c=\mathcal{L}_\xi \c=L_\xi \c+\xi^r\dot{\c}=\xi_o^i\pa_i \c+\co(r^{-2\x^2/3} e^{-2r/\ell}),
\eea
where $L_\xi$ denotes the Lie derivative w.r.t. the transverse components $\xi^i$ of the bulk vector field $\xi$.
It follows that the sources, $\bar g\sub{0}_{ij}(x)$, $\bar\l\sub{0}(x)$ and $\bar\c\sub{0}(x)$ transform under such a diffeomorphism as
\be\label{source-transformation}
\framebox[2.8in]{\rule[-0.7in]{0.0in}{0.0in}
\begin{minipage}{3.5in}
\begin{eqnarray*}
&&\d_\x \bar g\sub{0}_{ij}=L_{\x_o}\bar g\sub{0}_{ij}+\frac{2}{\ell}\d\s(x)\bar g\sub{0}_{ij},\NO\\
&&\d_\x \bar\l\sub{0}=\x_o^i(x)\pa_i\bar\l\sub{0}-\frac{1}{b_0 \ell}\d\s(x),\NO\\
&&\d_\x \bar\c\sub{0}= \x_o^i(x)\pa_i\bar\c\sub{0}.
\end{eqnarray*}
\end{minipage}
}
\ee
There is nothing surprising about the transformation of the metric and axion sources. They are exactly as they would be in
the case of strictly asymptotically AdS space. Namely, the bulk diffeomorphisms that preserve the form of the asymptotic expansions 
correspond to arbitrary boundary diffeomorphisms parameterized my $\x_o^i(x)$, as well as boundary Weyl transformations, parameterized
by the function $\s(x)$. What is rather unusual, is the transformation of the dilaton source, $\bar{\l}\sub{0}$ under 
the Weyl transformation $\d\s$. Contrary to the usual multiplicative transformation of the sources, the transformation of
$\bar{\l}\sub{0}$ is {\em additive} under boundary Weyl rescalings. This means that one can in fact remove the dilaton source completely
by means of a boundary Weyl rescaling. 

We can now also determine the transformation of the one-point functions under boundary Weyl rescalings. The above transformations 
of the sources imply that the corresponding functional derivatives transform as    
\be
\d_\s\left(\frac{\d}{\d\bar{g}\sub{0}_{ij}}\right)=-\frac{2}{\ell}\d\s(x)\frac{\d}{\d\bar{g}\sub{0}_{ij}},
\quad \d_\s\left(\frac{\d}{\d\bar{\l}\sub{0}}\right)=0,\quad \d_\s\left(\frac{\d}{\d\bar{\c}\sub{0}}\right)=0. 
\ee
Moreover, the renormalized action 
\be
\cs_{ren}=\lim_{r\to\infty}\widehat\cs\sub{4},
\ee
is invariant under {\em any} bulk diffeomorphism since the boundary term (\ref{IHQCD-counterterms}) does not 
break the bulk diffeomorphisms. Hence, 
\bea
&&\d_\s\left\langle T^i_j\right\rangle_{ren}=\d_\s\left(\frac{1}{\sqrt{\bar g\sub{0}}}\bar g\sub{0}^{ik}\frac{\d\cs_{ren}}{\d\bar g\sub{0}_{kj}}\right)=
-\frac{4}{\ell}\d\s\left\langle T^i_j\right\rangle_{ren},\NO\\
&&\d_\s\left\langle\co_\l\right\rangle_{ren}=\d_\s\left(\frac{1}{\sqrt{\bar g\sub{0}}}\frac{\d\cs_{ren}}{\d\bar\l\sub{0}}\right)
=-\frac{4}{\ell}\d\s\left\langle\co_\l\right\rangle_{ren},\NO\\
&&\d_\s\left\langle\co_\c\right\rangle_{ren}=\d_\s\left(\frac{1}{\sqrt{\bar g\sub{0}}}\frac{\d\cs_{ren}}{\d\bar\c\sub{0}}\right)
=-\frac{4}{\ell}\d\s\left\langle\co_\c\right\rangle_{ren},
\eea
that is all one-point functions transform {\em homogeneously} under boundary Weyl rescalings. 

The Ward identities now follow from the identity
\be
\d_\x\cs_{ren}=\int d^dx \left(-\frac12\d_\x\bar g\sub{0}_{ij}\left\langle T^{ij}\right\rangle_{ren}+
\d_\x\bar\l\sub{0}\left\langle\co_\l\right\rangle_{ren}+\d_\x\bar\c\sub{0}\left\langle\co_\c\right\rangle_{ren}\right)=0.
\ee
Inserting the transformation of the sources under the bulk diffeomorphisms considered above and using the fact that $\d\s(x)$ and 
$\x_o^i(x)$ are arbitrary leads respectively to   
\be\label{Ward-identities}
\framebox[10.5cm]{\rule[-1.5cm]{0cm}{0cm}
\begin{minipage}{15.0cm}
\begin{center}
\begin{eqnarray*}
&&\left\langle T^i_i\right\rangle_{ren}=-\frac{1}{b_0}\left\langle\co_\l\right\rangle_{ren},\NO\\\NO\\
&&D\sub{0}i\left\langle T^i_j\right\rangle_{ren}+\left\langle\co_\l\right\rangle_{ren}\pa_j\bar\l\sub{0}
+\left\langle\co_\c\right\rangle_{ren}\pa_j\bar\c\sub{0}=0.
\end{eqnarray*}
\end{center}
\end{minipage}
}
\ee
An immediate consequence of the trace Ward identity is that the tensor $\Om_{ij}$ introduced above is traceless. 

Finally, let us examine a bit closer the relation between the dilaton source $\bar\l\sub{0}$ and boundary Weyl transformations.
Note that the assignment of sources in the leading order asymptotic solutions (\ref{leading-asymptotics}) is rather arbitrary. 
In particular, we could have included a factor of the dilaton source in the definition of the boundary metric. Suppose, in particular,
that we define
\be\label{}
\framebox[5.7cm]{\rule[0cm]{0cm}{0cm}
\begin{minipage}{5.7cm}
\begin{center}
$\bar g\sub{0}_{ij}(x)=e^{-2b_0\bar{\l}\sub{0}(x)} \check g\sub{0}_{ij}(x),$
\end{center}
\end{minipage}
}
\ee 
Then, inserting this boundary metric back in (\ref{leading-asymptotics})\footnote{The same can be done with the higher order
terms in the asymptotic expansions but we will not do this explicitly here.} and evaluating the variation of the induced 
fields with respect to variations of the dilaton source $\bar{\l}\sub{0}$, one immediately sees from  
(\ref{asymptotic-diffeo-transformations}) that transformations of the dilaton source correspond precisely to boundary
Weyl transformations upon the identification 
\be\label{}
\framebox[3.7cm]{\rule[0cm]{0cm}{0cm}
\begin{minipage}{3.7cm}
\begin{center}
$\s(x)=-b_0\ell \bar{\l}\sub{0}(x).$
\end{center}
\end{minipage}
}
\ee
This observation proves that the source of the dilaton is gauge freedom that can be removed by a bulk diffeomorphism
corresponding to a boundary Weyl transformation. This property can also be used to restore the full dependence on the dilaton source 
of the one-point functions of the stress tensor and the axion.

\section{Concluding remarks}
\label{conclusions}
\setcounter{equation}{0}

We considered a generic dilaton-axion system coupled to Einstein-Hilbert gravity in arbitrary spacetime dimension and we 
carried out the procedure of holographic renormalization of this action for dimension up to and including five dimensions. 
The general boundary term that renders the variational problem for this action well defined is summarized in Tables 
\ref{order-1} and \ref{order-2}. This result is applicable to a very wide range of holographic models in the literature, including
$\cn=4$ super Yang-Mills in four dimensions, Improved Holographic QCD and non-conformal branes. We explicitly evaluated 
this general boundary term for a constant dilaton potential, corresponding to the standard dilaton-axion system 
dual to the complexified coupling of $\cn=4$ super Yang-Mills in four dimensions, in Appendix \ref{constant-potential}, 
and for IHQCD in Section \ref{IHQCD-application}. In particular, we systematically derived the generalized Fefferman-Graham
asymptotic expansions, provided exact expressions for the one-point functions in the presence of sources, and proved
the holographic holographic Ward identities by studying the asymptotic bulk diffeomorphisms that preserve the 
form of the asymptotic expansions. 

In the case of IHQCD, an important lesson from the analysis is that the source of the dilaton is not a physical coupling, but
its value can be thought of as an energy scale. In particular, changes in the dilaton source can be absorbed by a Weyl 
rescaling of the boundary metric. Moreover, the operator dual to the dilaton field $\l$ is the combination 
\be
\co_\l=\b(\l_{YM})\Tr F^2,
\ee 
which has fixed scaling dimension 4 under renormalization group flow. This is the combination that appears in the
trace Ward identity \ref{Ward-identities}, since the coefficient relating the trace of the stress tensor to the 
operator $\co_\l$ is a constant, independent of the renormalization group scale $\bar\l\sub{0}$.          

Our calculation for IHQCD, independently of the legitimacy of the model as a physically sound holographic model dual to 
pure Yang-Mills theory in four dimensions, provides us with an explicit example of a gravity model that can accommodate 
operators with running scaling dimensions. This is particularly interesting since it allows us, in principle, to 
develop supergravity holographic models capturing the dynamics of operators with non-protected scaling dimensions.

\section*{Acknowledgments}

We thank Elias Kiritsis, Umut Gursoy, Kostas Skenderis, George Papadopoulos, David Mateos and Diego Trancanelli for useful discussions and comments. 
We are especially grateful to David Mateos and Diego Trancanelli for carefully checking the results in Appendix 
\ref{constant-potential} and pointing out typos. We also thank the Galileo Galilei Institute for Theoretical Physics for the 
hospitality and the INFN for partial support during the completion of this work.

\appendix

\renewcommand{\thesection}{\Alph{section}}
\renewcommand{\theequation}{\Alph{section}.\arabic{equation}}

\section*{Appendix}
\setcounter{section}{0}

\section{Functional integration}
\label{functional-integration}
\setcounter{equation}{0}

In this appendix we outline the derivation of the functional integration formulas in Table \ref{f-integration}.
In particular, the question we want to address is the following: given a local functional $\car\sub{2n}(\vf)$ of the scalar
field, $\vf$, what is the local functional $F\sub{2n}(\vf)$ such that (cf. (\ref{integration-formula}))
\be\label{integration-formula-appendix}
\frac{\d\vf}{U'(\vf)}e^{a_n(\vf)}\car\sub{2n}(\vf)=\d_\vf F\sub{2n}(\vf)+e^{a_n(\vf)}\pa_i v\sub{2n}^i(\vf,\d\vf),
\ee
where $a_n(\vf)$ is a prescribed function of $\vf$ and $v\sub{2n}^i(\vf,\d\vf)$ is some vector field? If the source
$\car\sub{2n}(\vf)$ does not involve spacetime derivatives of the scalar field the answer to this question is given
by simple integration. However, if $\car\sub{2n}(\vf)$ involves derivatives of $\vf$, then determining
$F\sub{2n}(\vf)$ becomes less trivial. It turns out that one can still find a general formula if $\car\sub{2n}(\vf)$
involves first derivatives of the scalar field, but once $\car\sub{2n}(\vf)$ contains second and higher order derivatives
of the scalar field finding a general formula becomes much harder. What we will do instead here is to consider
only the sources $\car\sub{2n}(\vf)$ which are relevant to our computation of the solution of the Hamilton-Jacobi
equation in the main body of the paper.

\begin{itemize}
 \item $\car\sub{2n}(\vf)=r_{1^m}(\vf)t^{i_1i_2\ldots i_{m}}\pa_{i_1}\vf\pa_{i_2}\vf\ldots\pa_{i_m}\vf$ \\

 The first example is a generic source that is polynomial in first derivatives. Here, $t^{i_1i_2\ldots i_{m}}$
 is an arbitrary totally symmetric tensor that does not depend on $\vf$. In this case we can write
 $F\sub{2n}$ as
 \be
 F\sub{2n}=e^{a_n}\left(\a(\vf)t^{i_1i_2\ldots i_{m}}\pa_{i_1}\vf\pa_{i_2}\vf\ldots\pa_{i_m}\vf+
 D_i\left(\b(\vf)t^{ii_2\ldots i_{m}}\vf\pa_{i_2}\vf\ldots\pa_{i_m}\vf\right)\right),
 \ee
 where the functions $\a(\vf)$ and $\b(\vf)$ are to be determined. Evaluating the variation of
 this expression and inserting the result in (\ref{integration-formula-appendix}) we obtain the two equations
 \bea
 &&\b=\frac{m}{a'_n}\a\NO\\
 &&\a'+\left(a'_n-m\frac{a''_n}{a'_n}\right)\a=\frac{r_{1^m}}{U'},
 \eea
 which can be solved to determine $\a(\vf)$ and $\b(\vf)$. Since $\b(\vf)$ contributes a total
 derivative to Hamilton's principal function, i.e. to $e^{-a_n(\vf)}F\sub{2n}$, we are only interested in $\a(\vf)$,
 which is given by
 \be
 \a(\vf)=a_n'^m e^{-a_n}\int^\vf \frac{d\bar\vf}{U'}e^{a_n}a_n'^{-m} r_{1^m}(\bar\vf)=\fint_{n,m}^{\vf}r_{1^m}(\bar\vf).
 \ee

\item  $r_2(\vf)t^{ij}D_iD_j\vf$\\

Similarly, for a source with a single second derivative of the scalar field we can write
\be
F\sub{2n}=e^{a_n}\left(\a(\vf)t^{ij}D_iD_j\vf+\b(\vf)t^{ij}\pa_i\vf\pa_j\vf+D_i\left(\g(\vf)t^{ij}\pa_j\vf
+\d(\vf)D_jt^{ij}\right)\right).
\ee
Here, $t^{ij}$ is again a symmetric tensor that does not depend on $\vf$. Inserting the variation of this expression in 
(\ref{integration-formula-appendix}) leads to the following equations for the functions $\a(\vf)$, $\b(\vf)$, $\g(\vf)$ and $\d(\vf)$
\bea
&&a'_n(\b+\g')+\a''-\b'=0,\NO\\
&&a'_n(\g+\d')+2(\a'-\b)=0,\NO\\
&&a'_n\d+\a=0,\NO\\
&&2\a'+a'_n(\a+\g)-2\b=\frac{r_2}{U'}.
\eea
These can be immediately solved to obtain
\bea
&&\a(\vf)=a_n'e^{-a_n}\int^\vf \frac{d\bar\vf}{U'}e^{a_n}a_n'^{-1}r(\bar\vf)=\fint_{n,1}^\vf r(\bar\vf),\\
&&\b(\vf)=-a_n'^2e^{-a_n}\int^\vf \frac{d\bar\vf}{U'}e^{a_n}a_n'^{-2}U'a_n'\pa_{\bar{\vf}}^2\left(\frac{1}{a_n'}\right)\a(\bar\vf)
=-\fint_{n,2}^\vf U'a_n'\pa_{\bar{\vf}}^2\left(\frac{1}{a_n'}\right)\a(\bar\vf).\NO
\eea

\item $\left(r_{1^22}(\vf)t_1^{ijkl}+s_{1^22}(\vf)t_2^{ijkl}\right)\pa_i\vf\pa_j\vf D_kD_l\vf+
\left(r_{2^2}(\vf)t_1^{ijkl}+s_{2^2}(\vf)t_2^{ijkl}\right)D_iD_j\vf D_kD_l\vf$\\

As a final example we consider a generic term with four derivatives, but we restrict to covariantly constant
tensors $t_1^{ijkl}$ and $t_2^{ijkl}$. In particular, for our purposes it suffices to take these two
tensors to be the two linearly independent tensors constructed out of the metric. Namely, we will take
\be
t_1^{ijkl}=\frac13\left(\g^{ik}\g^{jl}+\g^{il}\g^{jk}+\g^{ij}\g^{kl}\right),\quad
t_2^{ijkl}=\frac13\left(\g^{ik}\g^{jl}+\g^{il}\g^{jk}-2\g^{ij}\g^{kl}\right).
\ee
Moreover, we need not consider a source for four first derivatives since we have already computed the result
for an arbitrary number of first derivatives above. Writing then
\bea
F\sub{2n}&=&e^{a_n}\left(A^{ijkl}D_iD_j\vf D_kD_l\vf +B^{ijkl}\pa_i\vf\pa_j\vf D_kD_l\vf +C^{ijkl}\pa_i\vf\pa_j\vf\pa_k\vf\pa_l\vf
\right.\NO\\
&&\left. +D_i\left(E^{ijkl}D_jD_kD_l\vf +H^{ijkl}\pa_j\vf D_kD_l\vf +G^{ijkl}\pa_j\vf\pa_k\vf\pa_l\vf\right)\right),
\eea
and inserting the variation of this expression in (\ref{integration-formula-appendix}) we obtain the set of coupled equations
\bea
&&3A'^{ijkl}+a'_n A^{ijkl}+B^{ilkj}+B^{ikjl}-2B^{ijkl}+a'_n H^{ijkl}=\frac{1}{U'}\left(r_{2^2}(\vf)t_1^{ijkl}+s_{2^2}(\vf)t_2^{ijkl}\right)\NO\\
&&2A'^{ijkl}+2\left(B'^{kjil}+B'^{ljik}\right)-12C^{ijkl}+a'_n\left(B^{ijkl}+H'^{ijkl}+G^{klij}+G^{ljik}+G^{kjil}\right)=\NO\\
\NO&&\frac{1}{U'}\left(r_{1^22}(\vf)t_1^{ijkl}+s_{1^22}(\vf)t_2^{ijkl}\right),\NO\\
&& -3C'^{ijkl}+B''^{(ijkl)}+a'_n\left(C^{ijkl}+G'^{(ijkl)}\right)=0,\NO\\
&&4 A'^{ijkl}+B^{kijl}+B^{lijk}-2B^{ijkl}+a'_n\left(E'^{ijkl}+H^{ijkl}\right)=0,\NO\\
&&2A^{ijkl}+a'_n E^{ijkl}=0.
\eea
Here, the parentheses in the indices mean total symmetrization. Note that these are five equations for six undetermined tensors.
They can be solved as follows. First we can use the last two equations to eliminate $E^{ijkl}$ and $H^{ijkl}$. This leads to the decoupled equation
\be
A'^{ijkl}+\left(a'_n-\frac{2a''_n}{a'_n}\right)A^{ijkl}=\frac{1}{U'}\left(r_{2^2}(\vf)t_1^{ijkl}+s_{2^2}(\vf)t_2^{ijkl}\right),
\ee
whose solution is
\be
A^{ijkl}=a_n'^2e^{-a_n}\int^\vf\frac{d\bar\vf}{U'}e^{a_n}a_n'^{-2}\left(r_{2^2}(\vf)t_1^{ijkl}+s_{2^2}(\vf)t_2^{ijkl}\right).
\ee
Next, we can set
\be
a'_nG^{ijkl}=4C^{ijkl},
\ee
so that we obtain the following two equations for $B^{ijkl}$ and $C^{ijkl}$:
\bea
&&2B'^{ijkl}+B'^{kijl}+B'^{lijk}+\left(a'_n-\frac{2a''_n}{a'_n}\right)B^{ijkl}+\frac{2a''_n}{a'_n}\left(B^{kijl}+B^{lijk}\right)\NO\\
&&=\frac{1}{U'}\left(r_{1^22}(\vf)t_1^{ijkl}+s_{1^22}(\vf)t_2^{ijkl}\right)-2a'_n\pa_\vf^2\left(\frac{1}{a_n'}\right)A^{ijkl},
\eea
and
\be
C'^{ijkl}+\left(a'_n-\frac{4a''_n}{a'_n}\right)C^{ijkl}+B''^{(ijkl)}=0.
\ee
The last equation for $C^{ijkl}$ gives
\be
C^{ijkl}=a_n'^4e^{-a_n}\int^\vf\frac{d\bar\vf}{U'}e^{a_n}a_n'^{-4}\left(-B''^{(ijkl)}\right).
\ee

To solve for $B^{ijkl}$ we first decompose it as
\be
B^{ijkl}=\a(\vf)t_{1}^{ijkl}+\b(\vf)t_2^{ijkl}.
\ee
Inserting this into the above equation for $B^{ijkl}$ leads to the two decoupled equations
\bea
&&4\a'+a'_n\a=\frac{1}{U'}\s,\NO\\
&&\b'+\left(a'_n-\frac{3a_n''}{a_n'}\right)\b=\frac{1}{U'}\om,
\eea
where
\bea
&&\s=r_{1^22}(\vf)-2U'a'_n\pa_\vf^2\left(\frac{1}{a_n'}\right)a_n'^2e^{-a_n}\int^\vf\frac{d\bar\vf}{U'}e^{a_n}a_n'^{-2}r_{2^2}(\bar\vf),\NO\\
&&\om=s_{1^22}(\vf)-2U'a'_n\pa_\vf^2\left(\frac{1}{a_n'}\right)a_n'^2e^{-a_n}\int^\vf\frac{d\bar\vf}{U'}e^{a_n}a_n'^{-2}s_{2^2}(\bar\vf).
\eea
These equations give
\bea
&&\a=\frac14 e^{-a_n/4}\int^\vf \frac{d\bar\vf}{U'}e^{a_n/4}\s,\NO\\
&&\b=a_n'^3e^{-a_n}\int^\vf \frac{d\bar\vf}{U'}e^{a_n}a_n'^{-3}\om.
\eea
Moreover, since $B^{(ijkl)}=\a t_1^{ijkl}$, we have
\be
C^{ijkl}=-t_1^{ijkl}a_n'^4e^{-a_n}\int^\vf\frac{d\bar\vf}{U'}e^{a_n}a_n'^{-4}\a''.
\ee

These results can be simplified considerably by noticing that the terms involving $\a$ combine into a total
derivative, up to a homogeneous term that is irrelevant since it contributes a finite piece. To see this, integrate
by parts the first term in
\be
\a t_1^{ijkl}\pa_i\vf\pa_j\vf D_kD_l\vf+ct_1^{ijkl}\pa_i\vf\pa_j\vf\pa_k\vf\pa_l\vf,
\ee
which gives
\be
-2\a t_1^{ijkl}\pa_i\vf\pa_j\vf D_kD_l\vf+(c-\a')t_1^{ijkl}\pa_i\vf\pa_j\vf\pa_k\vf\pa_l\vf.
\ee
We can now replace the first of these expressions in $F\sub{2n}$ by $2/3$ of the first expression plus
$1/3$ of the second, such that the coefficient of $t_1^{ijkl}\pa_i\vf\pa_j\vf D_kD_l\vf$ vanishes. This
replaces $c$ by $\tilde{c}=c-\a'/3$. However, we have seen above that the only source of the equation satisfied by
$c$, or $\tilde{c}$ now, is the coefficient of $t_1^{ijkl}\pa_i\vf\pa_j\vf D_kD_l\vf$. Since we have now set this coefficient
to zero, $\tilde{c}$ satisfies a homogeneous equation and hence we can set $\tilde{c}=0$. We therefore conclude that,
without loss of generality, we have
\be
B^{ijkl}=\b(\vf)t_2^{ijkl}, \quad C^{ijkl}=0,
\ee
for any $\s(\vf)$. Notice, in particular, that the source $r_{1^22}(\vf)$ gives no contribution at all.

\end{itemize}

\section{Dilaton-axion system with constant dilaton potential}
\label{constant-potential}
\setcounter{equation}{0}

In this appendix\footnote{I am grateful to David Mateos and Diego Trancanelli for checking the results of this appendix
and for pointing out typos in a preliminary version. I would also like to thank Wissam Chemissany, Henriette Elvang, Marios Hadjiantonis and Jelle Hartong for pointing out additional typos that were corrected in v3. Of course, I am solely responsible for any remaining typos.} we focus 
on the special case of a constant dilaton potential ($\ell=1$) 
\be
V(\vf)=d(d-1)=12.
\ee
This special case drastically simplifies the solution of both the zero order problem (\ref{zero-order-eq}) and the 
linear recursion equations (\ref{linear-eq}). In particular, the first order differential equations are replaced by 
{\em algebraic} equations. Namely, we have\footnote{In addition to the constant solution $U(\vf)=-(d-1)$ of 
(\ref{zero-order-eq-reduced}) there is additionally a one-parameter family of non-constant solutions given by 
\cite{Freedman:2003ax,Papadimitriou:2004rz}
\be
U(\vf)=-(d-1)\cosh\left(\sqrt{\frac{d}{d-1}}(\vf-\vf_o)\right),
\ee 
for some arbitrary constant $\vf_o$ and AdS asymptotics requires that $\vf\to \vf_o$ asymptotically. However, this solution only 
allows for a constant non-normalizable mode for the dilaton, i.e. $\vf_o$ cannot be a function of the transverse coordinates $x$
and hence it does not correspond to the most general asymptotics. In fact, a domain wall with such a `fake superpotential' 
describes a vacuum where the operator dual to the dilaton field has acquired a VEV \cite{Papadimitriou:2004rz}. Finally note
that a special feature of equation (\ref{zero-order-eq}) with constant scalar potential is that the solution describing
the most general asymptotics, i.e. $U(\vf)=-(d-1)$, is {\em isolated} in the space of solutions, while the one-parameter family of solutions describes 
special asymptotics. This is the reverse of what happens for non-constant scalar potentials. }
\bea
\label{zero-order-eq-simplified}
&&-\frac{d}{d-1}U^2+d(d-1)=0,\\
\label{linear-eq-simplified}
&&-\left(\frac{d-2n}{d-1}\right)U\cl\sub{2n}=\car\sub{2n},\quad n>0.
\eea
From these we immediately obtain
\bea
&&U=-(d-1),\NO\\
&&\cl\sub{2}=-\frac{1}{2(d-2)\k^2}\sqrt{\g}\left(R[\g]-\pa_i\vf\pa^i\vf-Z(\vf)\pa_i\chi\pa^i\chi\right),
\eea
where we have fixed the sign of $U$ by demanding that the solution leads, via the relation 
\be
\dot{\g}_{ij}\sim -\frac{2}{d-1}U\g_{ij},
\ee
to the correct asymptotics for the induced metric $\g_{ij}$.  

In order to compute $\cl\sub{4}$ we need to compute the momenta from $\cl\sub{2}$. We easily get
\bea
&&\p\sub{2}^{ij}=\frac{1}{2(d-2)\k^2}\sqrt{\g}\left(R^{ij}-\pa^i\vf\pa^j\vf-Z(\vf)\pa^i\chi\pa^j\chi
-\frac12\g^{ij}\left(R-\pa_k\vf\pa^k\vf-Z(\vf)\pa_k\chi\pa^k\chi\right)\right), \NO\\
&&\p_\vf\sub{2}=-\frac{1}{(d-2)\k^2}\sqrt{\g}\left(\square_\g\vf-\frac12Z'(\vf)\pa_i\c\pa^i\c\right),\NO\\
&&\p_\c\sub{2}=-\frac{1}{(d-2)\k^2}\sqrt{\g}\left(Z(\vf)\square_\g\c+Z'(\vf)\pa_i\vf\pa^i\c\right).
\eea
Hence, from (\ref{sources}), we evaluate
\bea
\car\sub{4}&=&-2\k^2\g^{-\frac12}\left(\p\sub{2}^i_j\p\sub{2}^j_i-\frac{1}{d-1}\p\sub{2}^2
+\frac14 \p_\vf\sub{2}^2+\frac14 Z^{-1}(\vf)\p_\chi\sub{2}^2\right)\NO\\
&=&-\frac{1}{2\k^2}\frac{1}{(d-2)^2}\sqrt{\g}\left(\rule{0cm}{0.5cm} R^{ij}R_{ij}+(\pa_i\vf\pa^i\vf)^2+Z^2(\vf)(\pa_i\c\pa^i\c)^2
-2R^{ij}\pa_i\vf\pa_j\vf\right.\NO\\
&&\left.-2Z(\vf)R^{ij}\pa_i\c\pa_j\c+2Z(\vf)(\pa_i\vf\pa^i\c)^2-\frac{d}{4(d-1)}(R-\pa_i\vf\pa^i\vf-Z(\vf)\pa_i\c\pa^i\c)^2
\right.\NO\\
&&\left.+\left(\square_\g\vf-\frac12Z'(\vf)\pa_i\c\pa^i\c\right)^2+Z(\vf)\left(\square_\g\c+\pa_\vf\log Z(\vf)\pa_i\vf\pa^i\c\right)^2\rule{0cm}{0.0cm}\right).
\eea
Now, the recursion relations tell us that $\cl\sub{4}$ is given by
\be
(d-4)\cl\sub{4}=\car\sub{4},
\ee
which is ill defined in $d=4$. This problem is of course well known and it is related to the breakdown of full diffeomorphism
invariance of the Hamilton-Jacobi functional, which in turn leads to the conformal anomaly of the dual conformal field theory
\cite{Henningson:1998gx}. It was shown in \cite{Papadimitriou:2004ap} that the Hamilton-Jacobi approach can be still applied 
in this case and reproduces the results of the Fefferman-Graham expansion \cite{FG} provided the radial cut-off is related to the 
deviation of $d$ from the desired value of $4$. In particular, we set
\be
r_0=\frac{1}{d-4},
\ee
and define $\cl\sub{4}|_{r_0}=-2r_0\tilde\cl\sub{4}|_{r_0}$, so that 
\be
\tilde{\cl}\sub{4}=-\frac12\car\sub{4}.
\ee
This then leads to the full counterterm action for $d=4$. Namely, dropping again the subscript $0$ from the radial cut-off,
\bea\label{counterterms}
S_{ct}&=&\frac{1}{\k^2}\int d^4x \sqrt{\g}\left(\rule{0cm}{0.7cm}3+\frac{1}{4}\left(R-\pa_i\vf\pa^i\vf-Z(\vf)\pa_i\chi\pa^i\chi\right)
\right.\NO\\
&&\left.-\log e^{-2r}\frac{1}{16}\left(\rule{0cm}{0.7cm} R^{ij}R_{ij}+(\pa_i\vf\pa^i\vf)^2+Z^2(\vf)(\pa_i\c\pa^i\c)^2
-2R^{ij}\pa_i\vf\pa_j\vf-2Z(\vf)R^{ij}\pa_i\c\pa_j\c\right.\right.\NO\\
&&\left.\left.+2Z(\vf)(\pa_i\vf\pa^i\c)^2-\frac13(R-\pa_i\vf\pa^i\vf-Z(\vf)\pa_i\c\pa^i\c)^2
\right.\right.\NO\\
&&\left.\left.+\left(\square_\g\vf-\frac12Z'(\vf)\pa_i\c\pa^i\c\right)^2+Z(\vf)\left(\square_\g\c+\pa_\vf\log Z(\vf)\pa_i\vf\pa^i\c\right)^2\rule{0cm}{0.0cm}\right)\right).
\eea
This expression does not quite agree with the one given in (26) of \cite{Nojiri:1999jj}, which we believe is incorrect.

\subsection{Asymptotic expansions}

The above recursive solution of the Hamilton-Jacobi equation, implies that the canonical momenta take the form
\bea
&&\p^{ij}=\p\sub{0}^{ij}+\p\sub{2}^{ij}+(-2r)\tilde{\p}\sub{4}^{ij}+\p\sub{4}^{ij}+\ldots,
\eea
and similarly for the dilaton and the axion, obtained by differentiating Hamilton's principal
function w.r.t. the corresponding induced field. The form of these expansions implies in turn that the induced fields 
admit the asymptotic expansions 
\bea
&&\g_{ij}=e^{2r}\left(g\sub{0}_{ij}+e^{-2r}g\sub{2}_{ij}+e^{-4r}\left(-2r h\sub{4}_{ij}+g\sub{4}_{ij}\right)+\ldots\right),\NO\\
&&\vf=\vf\sub{0}+e^{-2r}\vf\sub{2}+e^{-4r}\left(-2r \tilde{\vf}\sub{4}+\vf\sub{4}\right)+\ldots,\NO\\
&&\c=\c\sub{0}+e^{-2r}\c\sub{2}+e^{-4r}\left(-2r \tilde{\c}\sub{4}+\c\sub{4}\right)+\ldots,
\eea
where the dots denote subleading terms that are unambiguously determined in terms of the terms shown. These are precisely the 
well known Fefferman-Graham expansions \cite{FG}. 

The coefficients in these expansions can be easily deduced from the expressions for the canonical momenta we obtained in the
previous section. In particular, inserting the asymptotic expansions for the induced fields in the expressions 
(\ref{momenta-scalars}) for the canonical momenta on the one hand, and in the above expansions of the momenta on the other 
and comparing the two determines at second order
\bea
\vf\sub{2}&=&\frac{1}{4}\left(\square\sub{0}\vf\sub{0}-\frac12Z'(\vf\sub{0})g\sub{0}^{ij}\pa_i\c\sub{0}\pa_j\c\sub{0}\right),\NO\\
\c\sub{2}&=&\frac{1}{4}\left(\square\sub{0}\c\sub{0}+\frac{Z'(\vf\sub{0})}{Z(\vf\sub{0})}g\sub{0}^{ij}\pa_i\c\sub{0}\pa_j\vf\sub{0}\right),\NO\\
g\sub{2}_{ij}&=&-\frac12\left(R[g\sub{0}]_{ij}-\pa_i\vf\sub{0}\pa_j\vf\sub{0}-Z(\vf\sub{0})\pa_i\chi\sub{0}\pa_j\chi\sub{0}\right)\NO\\
&&+\frac{1}{12}g\sub{0}_{ij}\left(R[g\sub{0}]-\pa_k\vf\sub{0}\pa^k\vf\sub{0}-Z(\vf\sub{0})\pa_k\chi\sub{0}\pa^k\chi\sub{0}\right).
\eea
Comparing the logarithmic terms gives
\bea
&&-\frac{1}{\k^2}\left(h\sub{4}^{ij}-\Tr h\sub{4} g\sub{0}^{ij}\right)=\lim_{r\to\infty}\frac{1}{\sqrt{\g}}\tilde\p\sub{4}^{ij}
=\lim_{r\to\infty}\frac{1}{\sqrt{\g}}\frac{\d}{\d\g_{ij}}\int d^4x \tilde\cl\sub{4},\NO\\
&&-\frac{4}{\k^2}\tilde{\vf}\sub{4}=\lim_{r\to\infty}\frac{1}{\sqrt{\g}}\tilde\p_\vf\sub{4}
=\lim_{r\to\infty}\frac{1}{\sqrt{\g}}\frac{\d}{\d\vf}\int d^4x \tilde\cl\sub{4},\NO\\
&&-\frac{4}{\k^2}Z(\vf\sub{0})\tilde{\c}\sub{4}=\lim_{r\to\infty}\frac{1}{\sqrt{\g}}\tilde\p_\vf\sub{4}
=\lim_{r\to\infty}\frac{1}{\sqrt{\g}}\frac{\d}{\d\c}\int d^4x \tilde\cl\sub{4}, 
\eea
which can be easily evaluated explicitly using the above expression for $\tilde\cl\sub{4}$ but we will not need these expressions here. 
Finally, the $\co(e^{-4r})$ terms lead to the expressions for the renormalized one-point functions in the presence of sources, 
namely
\bea
\langle\co_\vf\rangle_{ren}=\frac{1}{\sqrt{g\sub{0}}}\p_\vf\sub{4}&=&-\frac{4}{\k^2}\vf\sub{4}-\frac{2}{\k^2}\tilde{\vf}\sub{4}
+\frac{1}{2\k^2}\left(-g\sub{2}^{ij}D\sub{0}_iD\sub{0}_j\vf\sub{0}\right.\NO\\
&&\left.-\left(D\sub{0}^ig\sub{2}_{ij}-\frac12D\sub{0}_j\Tr g\sub{2}\right)D\sub{0}^j\vf\sub{0}+\square\sub{0}\vf\sub{2}\right.\NO\\
&&\left. -\frac12Z''(\vf\sub{0})\vf\sub{2}g\sub{0}^{ij}\pa_i\c\sub{0}\pa_j\c\sub{0}+\frac12Z'(\vf\sub{0})g\sub{2}^{ij}\pa_i\c\sub{0}\pa_j\c\sub{0}\right.\NO\\
&&\left.-Z'(\vf\sub{0})g\sub{0}^{ij}\pa_i\c\sub{0}\pa_j\c\sub{2}\right),
\eea
\bea
\langle\co_\c\rangle_{ren}=\frac{1}{\sqrt{g\sub{0}}}\p_\c\sub{4}&=&-\frac{4}{\k^2}Z(\vf\sub{0})\c\sub{4}-\frac{2}{\k^2}Z(\vf\sub{0})\tilde{\c}\sub{4}
-\frac{2}{\k^2}Z'(\vf\sub{0})\vf\sub{2}\c\sub{2}\NO\\
&&+\frac{1}{2\k^2}\left(Z'(\vf\sub{0})\vf\sub{2}\square\sub{0}\c\sub{0}-Z(\vf\sub{0})g\sub{2}^{ij}D\sub{0}_iD\sub{0}_j\c\sub{0}\right.\NO\\
&&\left.-\left(D\sub{0}^ig\sub{2}_{ij}-\frac12D\sub{0}_j\Tr g\sub{2}\right)D\sub{0}^j\c\sub{0}+Z(\vf\sub{0})\square\sub{0}\c\sub{2}\right.\NO\\
&&\left.+Z''(\vf\sub{0})\vf\sub{2}\pa_i\c\sub{0}\pa^i\vf\sub{0}-Z'(\vf\sub{0})g\sub{2}^{ij}\pa_i\vf\sub{0}\pa_j\c\sub{0}\right.\NO\\
&&\left.+Z'(\vf\sub{0})\pa_i\c\sub{2}\pa^i\vf\sub{0}+Z'(\vf\sub{0})\pa_i\c\sub{0}\pa^i\vf\sub{2}\right),
\eea
\bea
\langle T_{ij}\rangle_{ren}=-\frac{2}{\sqrt{g\sub{0}}}\p\sub{4}_{ij}&=&\frac{2}{\k^2}\left(g\sub{4}_{ij}-\Tr g\sub{4} g\sub{0}_{ij}\right)+\frac{1}{\k^2}\left(h\sub{4}_{ij}-\Tr h\sub{4} g\sub{0}_{ij}\right)\NO\\
&&-\frac{1}{\k^2}\left(\Tr g\sub{2} g\sub{2}_{ij}-\Tr(g\sub{2}^2) g\sub{0}_{ij}\right)\NO\\
&&+\frac{1}{2\k^2}\left(D\sub{0}_kD\sub{0}_{(i}g\sub{2}^k_{j)}-\frac12\square\sub{0}g\sub{2}_{ij}-\frac12D\sub{0}_iD\sub{0}_j\Tr g\sub{2}\right.\NO\\
&&\left.-2\pa_{(i}\vf\sub{0}\pa_{j)}\vf\sub{2}-Z'(\vf\sub{0})\vf\sub{2}\pa_i\c\sub{0}\pa_j\c\sub{0}
-2Z(\vf\sub{0})\pa_{(i}\c\sub{0}\pa_{j)}\c\sub{2}\right.\NO\\
&&\left.-\frac12g\sub{2}_{ij}\left(R[g\sub{0}]-\pa_k\vf\sub{0}\pa^k\vf\sub{0}-Z(\vf\sub{0})\pa_k\chi\sub{0}\pa^k\chi\sub{0}\right)\right.\NO\\
&&\left.+\frac12g\sub{0}_{ij}g\sub{2}^{kl}\left(R[g\sub{0}]_{kl}-\pa_k\vf\sub{0}\pa_l\vf\sub{0}-Z(\vf\sub{0})\pa_k\chi\sub{0}\pa_l\chi\sub{0}\right)\right.\NO\\
&&\left.-\frac12g\sub{0}_{ij}\left(D\sub{0}_kD\sub{0}_lg\sub{2}^{kl}-\square\sub{0}\Tr g\sub{2}-2\pa_k\vf\sub{0}\pa^k\vf\sub{2}\right.\right.\NO\\
&&\left.\left.-Z'(\vf\sub{0})\vf\sub{2}\pa_k\c\sub{0}\pa^k\c\sub{0}-2Z(\vf\sub{0})\pa_k\c\sub{0}\pa^k\c\sub{2}\right)\right).
\eea
Note that in some places we have symmetrized indices with weight one, i.e. $(ij)\equiv \frac12(ij+ji)$. It can be easily 
checked that the expression for the one-point function of the stress tensor evaluated at zero dilaton and axion sources
completely agrees with the expression in (3.12) of \cite{deHaro:2000xn}.\footnote{In making the comparison one should keep in mind 
that $R_{ijkl}^{there}=-R_{ijkl}^{here}$. Moreover, we believe that there are two typos in (3.12) of \cite{deHaro:2000xn}. Namely,
in the third line of (3.12), $-\Tr g\sub{2} g\sub{2}_{ij}$ should read $\Tr g\sub{2} g\sub{2}_{ij}$, while in the beginning
of the fourth line $R_{ik}R^k_j$ should read $2R_{ik}R^k_j$.} 

The holographic Ward identities can now be derived as in the main body of the text in the case of IHQCD. Namely, the 
second of the equations in (\ref{HJ-equations}) holds order by order in the expansion of the Hamilton-Jacobi functional 
in eigenfunctions of the dilatation operator. In particular, we have 
\be
D\sub{0}^i\langle T_{ij}\rangle_{ren}+\langle\co_\vf\rangle_{ren}D\sub{0}_j\vf\sub{0}+\langle\co_\c\rangle_{ren}D\sub{0}_j\c\sub{0}=0.
\ee
Moreover, considering the transformation of the renormalized action under the asymptotic diffeomorphisms that leave 
the Fefferman-Graham expansions form-invariant leads to the trace Ward identity
\be
\langle T^i_i\rangle_{ren}=\ca(g\sub{0},\vf\sub{0},\c\sub{0}),
\ee
where
\bea
\ca(\g,\vf,\c)&=&\frac{1}{8\k^2}\left(\rule{0cm}{0.5cm} R^{ij}R_{ij}+(\pa_i\vf\pa^i\vf)^2+Z^2(\vf)(\pa_i\c\pa^i\c)^2
-2R^{ij}\pa_i\vf\pa_j\vf-2Z(\vf)R^{ij}\pa_i\c\pa_j\c\right.\NO\\
&&\left.+2Z(\vf)(\pa_i\vf\pa^i\c)^2-\frac{d}{4(d-1)}(R-\pa_i\vf\pa^i\vf-Z(\vf)\pa_i\c\pa^i\c)^2
\right.\NO\\
&&\left.+\left(\square_\g\vf-\frac12Z'(\vf)\pa_i\c\pa^i\c\right)^2+Z(\vf)\left(\square_\g\c+\pa_\vf\log Z(\vf)\pa_i\vf\pa^i\c\right)^2\rule{0cm}{0.5cm}\right),
\eea
is the conformal anomaly.

As a final comment, we observe that since 
\be
\tilde{\cl}\sub{4}=-\frac12\car\sub{4}=\frac12\sqrt{\g}\ca,
\ee
the terms involving $h\sub{4}_{ij}$, $\tilde{\vf}\sub{4}$ and $\tilde{\c}\sub{4}$ in the above expressions for
the renormalized one-point functions can be ignored since they are scheme dependent and they can be removed by adding the
finite counterterm
\be
-\frac14\int d^4x \sqrt{\g}\ca.
\ee

\providecommand{\href}[2]{#2}\begingroup\raggedright\endgroup


\begin{thebibliography}{10}

\bibitem{Klebanov:2000hb}
I.~R. Klebanov and M.~J. Strassler, {\it {Supergravity and a confining gauge
  theory: Duality cascades and chi SB resolution of naked singularities}},
  {\em JHEP} {\bf 0008} (2000) 052,
  [\href{http://xxx.lanl.gov/abs/hep-th/0007191}{{\tt hep-th/0007191}}].

\bibitem{Maldacena:2000yy}
J.~M. Maldacena and C.~Nunez, {\it {Towards the large N limit of pure N=1
  superYang-Mills}},  {\em Phys.Rev.Lett.} {\bf 86} (2001) 588--591,
  [\href{http://xxx.lanl.gov/abs/hep-th/0008001}{{\tt hep-th/0008001}}].

\bibitem{Wiseman:2008qa}
T.~Wiseman and B.~Withers, {\it {Holographic renormalization for coincident
  Dp-branes}},  {\em JHEP} {\bf 0810} (2008) 037,
  [\href{http://xxx.lanl.gov/abs/0807.0755}{{\tt 0807.0755}}].

\bibitem{Kanitscheider:2008kd}
I.~Kanitscheider, K.~Skenderis, and M.~Taylor, {\it {Precision holography for
  non-conformal branes}},  {\em JHEP} {\bf 0809} (2008) 094,
  [\href{http://xxx.lanl.gov/abs/0807.3324}{{\tt 0807.3324}}]. * Temporary
  entry *.

\bibitem{Guica:2010sw}
M.~Guica, K.~Skenderis, M.~Taylor, and B.~C. van Rees, {\it {Holography for
  Schrodinger backgrounds}},  {\em JHEP} {\bf 1102} (2011) 056,
  [\href{http://xxx.lanl.gov/abs/1008.1991}{{\tt 1008.1991}}].

\bibitem{FG}
C.~Fefferman and C.~R. Graham, {\it {Conformal Invariants}},  {\em Elie Cartan
  et les Math{\'e}matiques d'aujourd'hui, Ast{\'e}risque} (1985).

\bibitem{Papadimitriou:2010as}
I.~Papadimitriou, {\it {Holographic renormalization as a canonical
  transformation}},  \href{http://xxx.lanl.gov/abs/1007.4592}{{\tt 1007.4592}}.

\bibitem{Papadimitriou:2005ii}
I.~Papadimitriou and K.~Skenderis, {\it {Thermodynamics of asymptotically
  locally AdS spacetimes}},  {\em JHEP} {\bf 08} (2005) 004,
  [\href{http://xxx.lanl.gov/abs/hep-th/0505190}{{\tt hep-th/0505190}}].

\bibitem{ElShowk:2011ag}
S.~El-Showk and K.~Papadodimas, {\it {Emergent Spacetime and Holographic
  CFTs}},  \href{http://xxx.lanl.gov/abs/1101.4163}{{\tt 1101.4163}}.

\bibitem{Gursoy:2007cb}
U.~Gursoy and E.~Kiritsis, {\it {Exploring improved holographic theories for
  QCD: Part I}},  {\em JHEP} {\bf 02} (2008) 032,
  [\href{http://xxx.lanl.gov/abs/0707.1324}{{\tt 0707.1324}}].

\bibitem{Gursoy:2010fj}
U.~Gursoy, E.~Kiritsis, L.~Mazzanti, G.~Michalogiorgakis, and F.~Nitti, {\it
  {Improved Holographic QCD}},  \href{http://xxx.lanl.gov/abs/1006.5461}{{\tt
  1006.5461}}.

\bibitem{Gibbons:1976ue}
G.~Gibbons and S.~Hawking, {\it {Action Integrals and Partition Functions in
  Quantum Gravity}},  {\em Phys.Rev.} {\bf D15} (1977) 2752--2756.

\bibitem{Mateos:2011ix}
D.~Mateos and D.~Trancanelli, {\it {The anisotropic N=4 super Yang-Mills plasma
  and its instabilities}},  \href{http://xxx.lanl.gov/abs/1105.3472}{{\tt
  1105.3472}}.

\bibitem{Mateos:2011tv}
D.~Mateos and D.~Trancanelli, {\it {Thermodynamics and Instabilities of a
  Strongly Coupled Anisotropic Plasma}},
  \href{http://xxx.lanl.gov/abs/1106.1637}{{\tt 1106.1637}}. * Temporary entry
  *.

\bibitem{Papadimitriou:non-conformal-branes}
I.~Papadimitriou, {\it Holographic renormalization of extended objects}. In preparation.

\bibitem{de_Boer:1999xf}
J.~de~Boer, E.~P. Verlinde, and H.~L. Verlinde, {\it {On the holographic
  renormalization group}},  {\em JHEP} {\bf 08} (2000) 003,
  [\href{http://xxx.lanl.gov/abs/hep-th/9912012}{{\tt hep-th/9912012}}].

\bibitem{Papadimitriou:2004ap}
I.~Papadimitriou and K.~Skenderis, {\it {AdS / CFT correspondence and
  geometry}},  \href{http://xxx.lanl.gov/abs/hep-th/0404176}{{\tt
  hep-th/0404176}}.

\bibitem{Henningson:1998gx}
M.~Henningson and K.~Skenderis, {\it {The holographic Weyl anomaly}},  {\em
  JHEP} {\bf 07} (1998) 023,
  [\href{http://xxx.lanl.gov/abs/hep-th/9806087}{{\tt hep-th/9806087}}].

\bibitem{Balasubramanian:1999re}
V.~Balasubramanian and P.~Kraus, {\it {A stress tensor for anti-de Sitter
  gravity}},  {\em Commun. Math. Phys.} {\bf 208} (1999) 413--428,
  [\href{http://xxx.lanl.gov/abs/hep-th/9902121}{{\tt hep-th/9902121}}].

\bibitem{Emparan:1999pm}
R.~Emparan, C.~V. Johnson, and R.~C. Myers, {\it {Surface terms as counterterms
  in the AdS/CFT correspondence}},  {\em Phys. Rev.} {\bf D60} (1999) 104001,
  [\href{http://xxx.lanl.gov/abs/hep-th/9903238}{{\tt hep-th/9903238}}].

\bibitem{Kraus:1999di}
P.~Kraus, F.~Larsen, and R.~Siebelink, {\it {The gravitational action in
  asymptotically AdS and flat spacetimes}},  {\em Nucl. Phys.} {\bf B563}
  (1999) 259--278, [\href{http://xxx.lanl.gov/abs/hep-th/9906127}{{\tt
  hep-th/9906127}}].

\bibitem{de_Haro:2000xn}
S.~de~Haro, S.~N. Solodukhin, and K.~Skenderis, {\it {Holographic
  reconstruction of spacetime and renormalization in the AdS/CFT
  correspondence}},  {\em Commun. Math. Phys.} {\bf 217} (2001) 595--622,
  [\href{http://xxx.lanl.gov/abs/hep-th/0002230}{{\tt hep-th/0002230}}].

\bibitem{Bianchi:2001kw}
M.~Bianchi, D.~Z. Freedman, and K.~Skenderis, {\it {Holographic
  Renormalization}},  {\em Nucl. Phys.} {\bf B631} (2002) 159--194,
  [\href{http://xxx.lanl.gov/abs/hep-th/0112119}{{\tt hep-th/0112119}}].

\bibitem{Martelli:2002sp}
D.~Martelli and W.~Mueck, {\it {Holographic renormalization and Ward identities
  with the Hamilton-Jacobi method}},  {\em Nucl. Phys.} {\bf B654} (2003)
  248--276, [\href{http://xxx.lanl.gov/abs/hep-th/0205061}{{\tt
  hep-th/0205061}}].

\bibitem{Arnowitt:1960es}
R.~L. Arnowitt, S.~Deser, and C.~W. Misner, {\it {Canonical variables for
  general relativity}},  {\em Phys.Rev.} {\bf 117} (1960) 1595--1602.

\bibitem{Lee:1990nz}
J.~Lee and R.~M. Wald, {\it {Local symmetries and constraints}},  {\em J. Math.
  Phys.} {\bf 31} (1990) 725--743.

\bibitem{Parry:1993mw}
J.~Parry, D.~S. Salopek, and J.~M. Stewart, {\it {Solving the Hamilton-Jacobi
  equation for general relativity}},  {\em Phys. Rev.} {\bf D49} (1994)
  2872--2881, [\href{http://xxx.lanl.gov/abs/gr-qc/9310020}{{\tt
  gr-qc/9310020}}].

\bibitem{Panero:2009tv}
  M.~Panero,
  {\it Thermodynamics of the QCD plasma and the large-N limit},
  {\em Phys.\ Rev.\ Lett.\ }  {\bf 103} (2009) 232001,
  [\href{http://xxx.lanl.gov/abs/0907.3719}{{\tt 0907.3719 [hep-lat]}}].
  

\bibitem{Papadimitriou:2006dr}
I.~Papadimitriou, {\it {Non-supersymmetric membrane flows from fake
  supergravity and multi-trace deformations}},  {\em JHEP} {\bf 02} (2007) 008,
  [\href{http://xxx.lanl.gov/abs/hep-th/0606038}{{\tt hep-th/0606038}}].

\bibitem{Freedman:2003ax}
D.~Z. Freedman, C.~Nunez, M.~Schnabl, and K.~Skenderis, {\it {Fake Supergravity
  and Domain Wall Stability}},  {\em Phys. Rev.} {\bf D69} (2004) 104027,
  [\href{http://xxx.lanl.gov/abs/hep-th/0312055}{{\tt hep-th/0312055}}].

\bibitem{Papadimitriou:2004rz}
I.~Papadimitriou and K.~Skenderis, {\it {Correlation functions in holographic
  RG flows}},  {\em JHEP} {\bf 0410} (2004) 075,
  [\href{http://xxx.lanl.gov/abs/hep-th/0407071}{{\tt hep-th/0407071}}].

\bibitem{Nojiri:1999jj}
  S.~Nojiri, S.~D.~Odintsov, S.~Ogushi, A.~Sugamoto and M.~Yamamoto,
  {\it Axion - dilatonic conformal anomaly from AdS / CFT correspondence},
  {\em Phys.\ Lett.\  B} {\bf 465} (1999) 128,
  [\href{http://xxx.lanl.gov/abs/hep-th/9908066}{{\tt hep-th/9908066}}].

\bibitem{deHaro:2000xn}
S.~de~Haro, S.~N. Solodukhin, and K.~Skenderis, {\it {Holographic
  reconstruction of spacetime and renormalization in the AdS/CFT
  correspondence}},  {\em Commun. Math. Phys.} {\bf 217} (2001) 595--622,
  [\href{http://xxx.lanl.gov/abs/hep-th/0002230}{{\tt hep-th/0002230}}].

\end{thebibliography}
\end{document}